\documentclass[intlimits,twoside,a4paper]{article}
\usepackage[cp1251]{inputenc}
\usepackage[eqsecnum]{cmpj3}

\usepackage{natbib}
\usepackage{amssymb} 
\usepackage{graphicx}
\usepackage{bm}

\def\state#1{\mid#1\rangle}

\def\astate#1{\langle#1\mid}

\def\ferrostate{\state{\uparrow \ldots \uparrow}}

\issue{2020}{23}{4}{43712}
\doinumber{10.5488/CMP.23.43712}

\title[Loop-gas description of the localized-magnon states on the kagome lattice with open boundary conditions]{Loop-gas description of the localized-magnon states on the kagome lattice with open boundary conditions}
\date{Received July 20, 2020, in final form August 29, 2020}

\author[A. Honecker, J. Richter, J. Schnack, A. Wietek]{A. Honecker\refaddr{adr:cergy},
J. Richter\refaddr{adr:magdeburg,adr:dresden},
J. Schnack\refaddr{adr:bielefld},
A. Wietek\refaddr{adr:flatiron}
}

\addresses{
\addr{adr:cergy}
Laboratoire de Physique Th\'eorique et Mod\'elisation, CNRS UMR 8089,
CY Cergy Paris Universit\'e, 95302 Cergy-Pontoise Cedex, France
\addr{adr:magdeburg}
Institut f\"ur Physik, Universit\"at Magdeburg, P.O. Box 4120, 39016 Magdeburg, Germany
\addr{adr:dresden}
Max-Planck-Institut f\"{u}r Physik Komplexer Systeme,
        N\"{o}thnitzer Stra{\ss}e 38, 01187 Dresden, Germany
\addr{adr:bielefld}
Fakult\"at f\"ur Physik, Universit\"at Bielefeld, Postfach 100131, 33501 Bielefeld, Germany
\addr{adr:flatiron}
Center for Computational Quantum Physics, Flatiron Institute,
New York, NY 10010, USA
}

\date{Received July 20, 2020, in final form August 29, 2020}
\begin{document}

\maketitle

\begin{abstract}
The high-field regime of the spin-$s$ XXZ antiferromagnet on the kagome
lattice gives rise to macroscopically degenerate ground states thanks
to a completely flat lowest single-magnon band. The corresponding
excitations can be localized on loops in real space and have been coined ``localized magnons''. Thus, the description of the many-body ground states
amounts  to characterizing the allowed classical loop configurations
and eliminating the quantum mechanical linear relations between them.
Here, we investigate this loop-gas description on finite kagome lattices
with open boundary conditions and compare the results with exact
diagonalization for the spin-1/2 XY model on the same lattice.
We find that the loop gas provides an exact account of the degenerate
ground-state manifold while a hard-hexagon description misses
contributions from nested loop
configurations. The densest packing
of the loops corresponds to a magnon crystal that according to
the zero-temperature magnetization curve is a stable ground
state of the spin-1/2 XY model in a window of magnetic fields of
about $4\%$ of the saturation field just below this saturation field.
We also present numerical results for the specific heat
obtained by the related methods of thermal pure quantum (TPQ) states
and the finite-temperature Lanczos method (FTLM). For a field
in the stability range of the magnon crystal, one finds a low-temperature
maximum of the specific heat that corresponds to a finite-temperature
phase transition into the magnon crystal at low temperatures.

\keywords
frustrated magnetism, Kagome lattice, XY model, lattice gases, phase transitions, exact diagonalization
\end{abstract}

\section{Introduction}

The study of the kagome lattice in condensed matter physics goes back at 
least to Sy\^ozi's famous investigation of the Ising model on this 
lattice \cite{Syozi51}. The kagome lattice is built from corner-sharing 
triangles. On the one hand, antiferromagnetic interactions along the 
three edges of each triangle compete so that not all of them can be 
satisfied at the same time. On the other hand, the connectivity of the 
corner-sharing arrangement is low so that the number of constraints 
arising from the coupling between triangles is also low (see, e.g., 
reference~\cite{MC98}). This entails a huge degeneracy of the Ising 
antiferromagnet on the kagome lattice \cite{KN53} and thus also preempts 
any finite-temperature phase transition in this case~\cite{Syozi51}. 
Investigation of the spin-1/2 Heisenberg antiferromagnet on the kagome 
lattice is a comparably young endeavor \cite{Elser89,ZE90,CE92,LE93}. 
Even if the classical ground-state degeneracy is lifted by quantum 
fluctuations, the nature of the ground state of the spin-1/2 kagome 
antiferromagnet has remained controversial in spite of numerous and big 
efforts, see 
references~\cite{Capponi04,SH07,EV10,Yan11,Depenbrock12,Depenbrock12ol,Liao17,He17,Mei17,Jiang19} 
for some examples. One distinguishing feature of the spin-1/2 kagome 
antiferromagnet is an exceptional number of low-lying singlet excitations 
on finite lattices \cite{Lecheminant97,Waldtmann98,Mila98,Laeuchli19} which 
renders the determination of the true ground state so delicate and 
challenging that some of the founding fathers of the field 
\cite{Elser89,ZE90,LE93} have recently suggested to suspend this endeavor 
\cite{Yao2020}.

The spin-1/2 kagome antiferromagnet in an applied magnetic field is also 
interesting since it exhibits several plateaux in its zero-temperature 
magnetization curve 
\cite{Hid:JPSJ01,CGH:PRB02,HSR:JP04,CGH:PRB05,NSH:NC13,CDH:PRB13,NaS:JPSJ18,ONO:NC19}. 
For some of these plateaux, the inclusion of the external magnetic field 
simplifies the situation in the sense that it allows for a rather 
unambiguous identification of the plateau states as valence-bond 
crystal-type states, at least for the plateaux at $1/3$, $5/9$, and $7/9$ 
of the saturation magnetization 
\cite{CGH:PRB05,NSH:NC13,CDH:PRB13,ONO:NC19}. Another intriguing 
phenomenon arises at the saturation field where part of the classical 
degeneracy survives at the quantum level, thus violating the third law of 
thermodynamics. The essence is a flat single-magnon band that permits a 
strict localization of the corresponding excitations in real space and 
thus permits the construction of exact many-body eigenstates 
\cite{SHS:PRL02,RDS:PRL04,ZhT:PRB04,ZhT:PTPS05,DeR:PRB04,DeR:EPJB06,DRH:LTP07}. 
These so-called ``localized-magnon states'' live on closed loops on the 
kagome lattice and can be shown to be not only exact eigenstates but in 
fact ground states in the corresponding sectors of the magnetic quantum 
number $S^z$ \cite{SSRS1,Schmidt2002}. In particular, the three-fold 
degenerate crystalline state of the $7/9$ plateau can be written down 
exactly \cite{SHS:PRL02}. More generally, the subset of the 
localized-magnon states corresponding to loops of minimum size can be 
mapped \cite{ZhT:PRB04,ZhT:PTPS05} to a model of hard hexagons that was 
solved exactly by Baxter \cite{Baxter1980,BaxterTsang1980}. However, 
there are nested loop configurations \cite{ZhT:PRB04,ZhT:PTPS05} that can 
be argued to give rise to a further macroscopic contribution of the 
ground-state degeneracy \cite{DRH:LTP07}.

Recently, we have investigated the full loop-gas description of the 
localized-magnon states on finite lattices with $N\leqslant 72$ sites 
\cite{SSHR20}. Following the general philosophy that periodic boundary 
conditions reduce finite-size effects by eliminating explicit boundaries, 
we  employed periodic boundary conditions in that investigation. The 
resulting non-trivial topology of a torus leads to loop configurations 
that wind around the boundaries. There are several issues associated
with these winding loops. Firstly, they lead to an enormous number of 
geometrically allowed configurations before taking linear relations into 
account~\cite{SSHR20}. Secondly, on a torus a nested configuration of two 
loops can actually be expressed as a linear combination of hard-hexagon 
and winding configurations, as is evidenced by the counting of ground 
states in the two-magnon sector \cite{DRH:LTP07,SSHR20}. Thirdly, one 
finds that the number of linearly independent loop configurations does 
not describe all ground states of the spin-1/2 kagome Heisenberg 
antiferromagnet on a torus \cite{SSHR20} which raises the question 
whether the loop gas does yield a complete description of the 
ground-state manifold of the spin-1/2 kagome Heisenberg antiferromagnet 
around the saturation field in the thermodynamic limit. In particular, the 
latter question urges us to investigate here the loop gas on finite 
systems with open boundary conditions. In order to ensure that the open 
boundary conditions do not spoil the localized-magnon states, it is 
advisable to consider the XY rather than the Heisenberg model in this 
case\footnote{Here, we follow the reference \cite{Derzhko02} and refer to
the $\Delta=0$ limit of the XXZ model as ``XY model'', while the $\Delta=1$ 
case would be the Heisenberg model.
}. In what follows, we first explain and present the results for the 
kagome loop gas with open boundary conditions. Then, we analyze the 
resulting thermodynamic properties and present a comparison with 
``exact'' diagonalization results for the spin-1/2 XY model.

\section{Model and localized-magnon states}

\label{sec:modelLoc}

Let us start by recalling some results for the spin-$s$ XXZ model in a magnetic
field \cite{RSH04}, given by the Hamiltonian
\begin{eqnarray}
H &=& J\,\sum_{\langle i,j \rangle} \left(S_i^x \, S_j^x + S_i^y \, S_j^y + \Delta\, S_i^z \, S_j^z\right)
  - h \, \sum_i S_i^z\nonumber\\
&=& J\,\sum_{\langle i,j \rangle} \left[\frac12\,\left(S_i^{+} \, S_j^{-} + S_i^{-} \, S_j^{+} \right) + \Delta\, S_i^z \, S_j^z\right]
  - h \, \sum_i S_i^z \, ,
\label{eq:H}
\end{eqnarray}
where $\langle i,j \rangle$ denotes the nearest-neighbor pairs of sites on
a kagome lattice, and the $\vec{S}_i$ are spin-$s$ operators at site $i$.
$J>0$ is an antiferromagnetic coupling constant and $\Delta$ is the exchange anisotropy.
We note in passing that the ground-state problem corresponding to $h=0$ in
equation~(\ref{eq:H}) was studied, e.g., in references~\cite{Goetze2015,He2015,Zhu2015},
and that the case of the one third-plateau was investigated for $s=1/2$
in reference~\cite{CGH:PRB05}.

\begin{figure}[tb!]
\centering
\includegraphics[width=9 true cm]{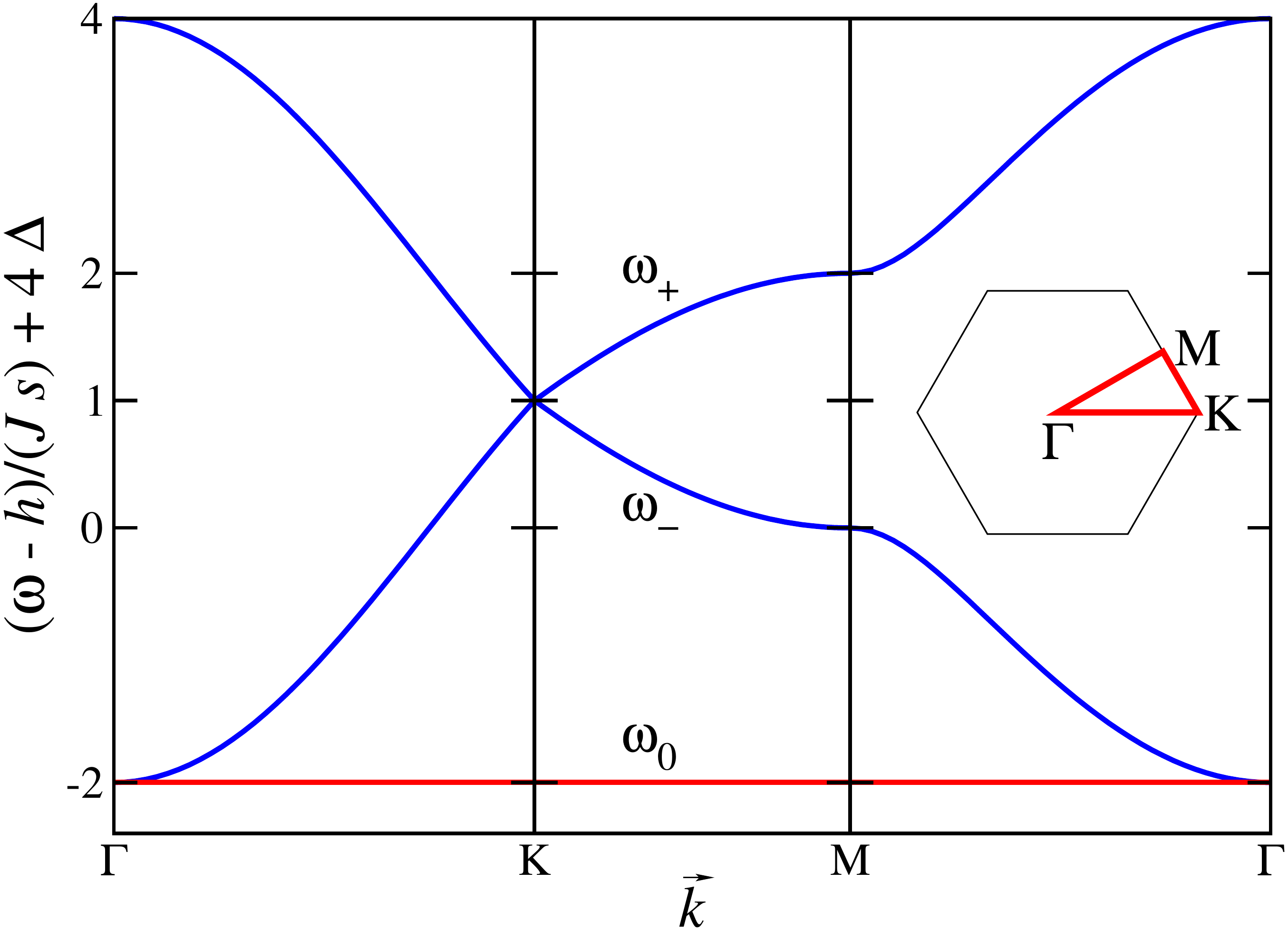}
\caption{(Colour online) The three bands $\omega_i(\vec{k})$ of single-magnon
energies on the kagome lattice along the path
in the Brillouin zone shown in the inset.
Note that $\omega_0(\vec{k})$ is completely independent of $\vec{k}$.}
\label{fig:Kenergy}
\end{figure}

A first simple observation is that the ferromagnetic state $\ferrostate$ 
is a trivial eigenstate of the Hamiltonian (\ref{eq:H}). If we consider 
the subspace where a single spin is flipped relative to this ferromagnetic 
state, we arrive at a single-particle problem that we  call 
``single-magnon'' problem. This single-magnon problem can still be solved 
in a closed form, at least in the thermodynamic limit. Let us impose 
periodic boundary conditions so that we can first transform the problem 
to the reciprocal space. Since a unit cell of the kagome lattice contains 
three sites, the one-magnon problem for the Hamiltonian (\ref{eq:H}) 
leads to a $3 \times 3$ matrix in reciprocal space whose eigenvalues 
yield three bands of single-magnon energies~\cite{RSH04}
\begin{eqnarray}
\omega_{0}(k_x,k_y) &=& h-J\,s\,(2+4\,\Delta) \, , \nonumber \\
\omega_{\pm}(k_x,k_y) &=& h + J\,s\,\left(1 \pm 1 \, \sqrt{
1 + 4 \, \cos\left({k_x \over 2}\right) \, \left[
\cos\left({\sqrt{3} \, k_y \over 2}\right) + \cos\left({k_x \over 2}\right)
 \right]} - 4 \,\Delta\right) \, .
\label{eq:singleMagnon}
\end{eqnarray}
Figure \ref{fig:Kenergy} shows these three single-magnon bands along
the path in the Brillouin zone sketched in the inset of this figure.
Note that the lowest band $\omega_0(\vec{k})$ turns out to be completely flat,
and this will be the cornerstone of the following discussion.

As a byproduct we can obtain the saturation field $h_{\rm sat}$
from equation~(\ref{eq:singleMagnon}). If the lowest excitation energy
$\omega_0$ becomes negative, then the ferromagnetic state $\ferrostate$ is
unstable. The condition $\omega_0(h_{\rm sat}) = 0$ yields
\begin{equation}
h_{\rm sat} = J\,s\,(2+4\,\Delta) \, .
\label{eq:Hsat}
\end{equation}
We  focus on the proximity of this saturation field in what follows.

\begin{figure}[!t]
\centering
\includegraphics[height=0.21\columnwidth]{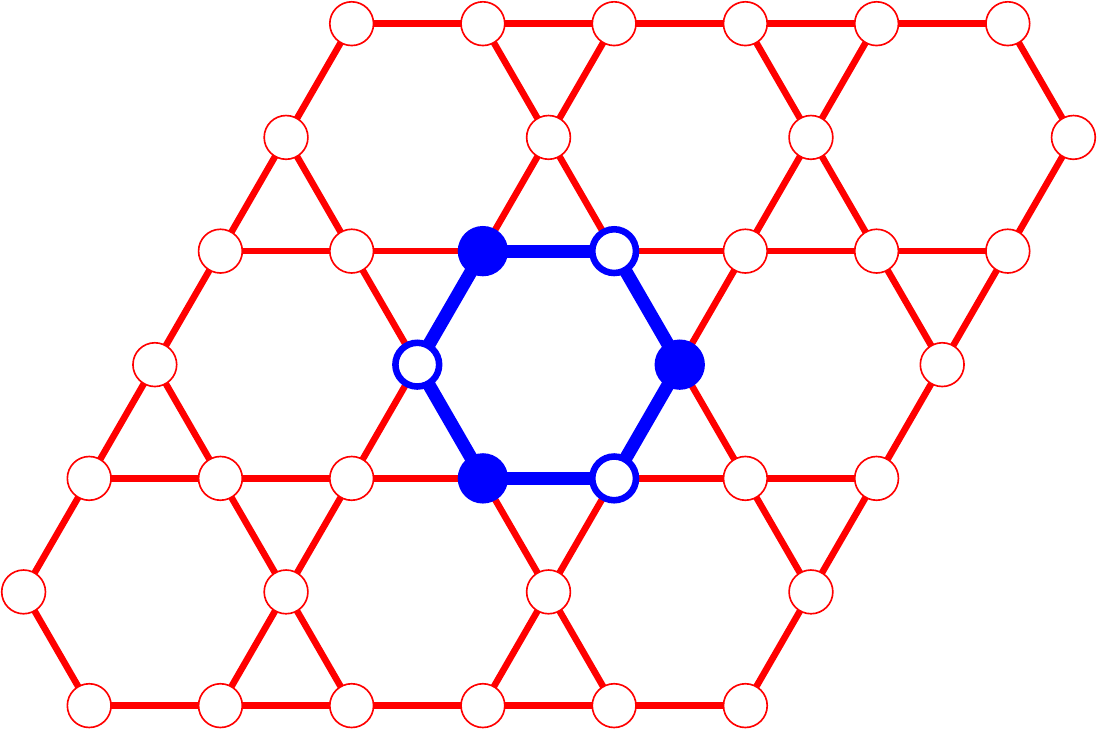}\quad%
\includegraphics[height=0.21\columnwidth]{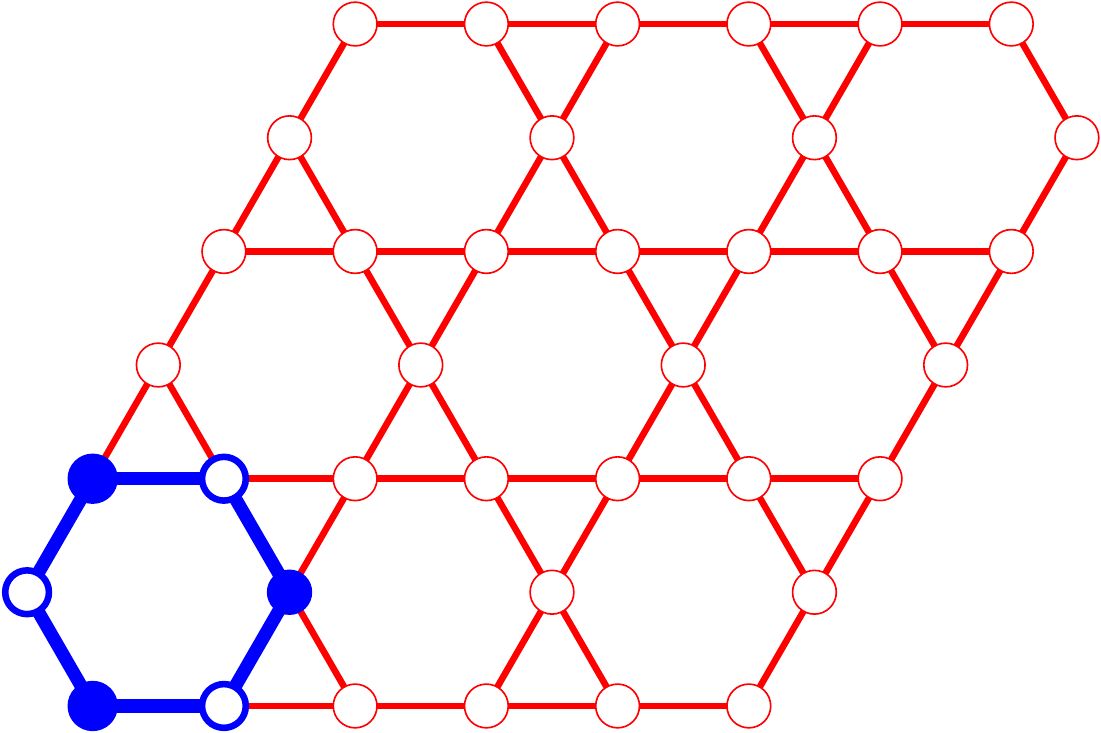}\quad%
\includegraphics[height=0.21\columnwidth]{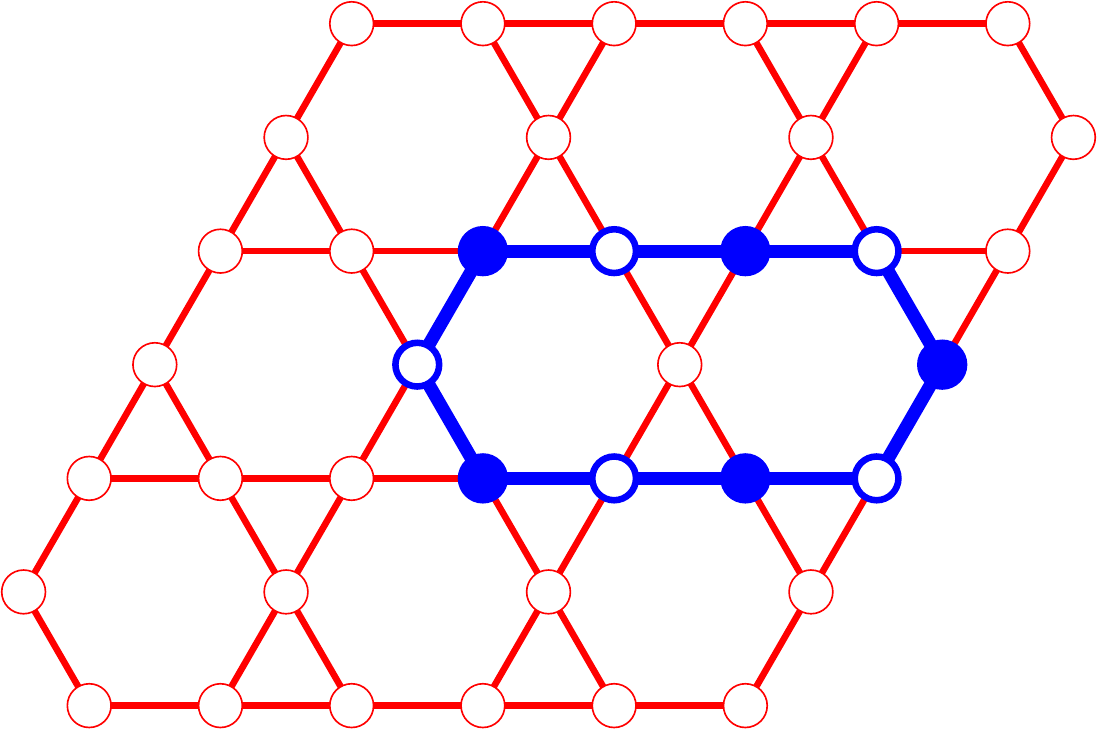}
\caption{(Colour online) Three one-loop configuration on an $N=38$ kagome lattice.
Sites and bonds belonging to a loop $\ell$ are shown in blue
(bold lines for the bonds). The filled and open circles correspond
to the alternating sign along the sites of the loop in equation~(\ref{eq:1loopState}).
The loop on the left is localized in the center and gives rise
to an exact eigenstate of the spin-$s$ XXZ model. The two loops
in the middle and at the right touch the boundary of the lattice such that
they yield exact eigenstates only in the XY limit $\Delta = 0$.
\label{fig:1loop}
}
\end{figure}

Let us return to the completely flat single-magnon branch $\omega_0$ of 
equation~(\ref{eq:singleMagnon}). Since it is independent of $\vec{k}$, one 
can construct linear combinations of the corresponding states that are 
completely localized. In fact, these localized excitations can be 
constructed in real space also for finite lattices with open and not just 
periodic boundary conditions. These localized single-magnon states have 
the form
\begin{equation}
\state{\ell} = 
\sum_{x\in \ell} (-1)^{x} \, S_{x}^- \, \ferrostate \, ,
\label{eq:1loopState}
\end{equation}
where $\ell$ is a closed loop of even length so
that it is always separated by at least one empty site when it bends
back onto itself. Some examples of such loops for an open lattice
with $N=38$ sites are shown in figure~\ref{fig:1loop}.

Two conditions ensure that the states equation~(\ref{eq:1loopState})
are exact eigenstates of the Hamiltonian (\ref{eq:H})~\cite{SHS:PRL02,Richter2004}:
\begin{enumerate}
\item The states equation~(\ref{eq:1loopState}) are exact eigenstates of the XY part
$\frac{J}{2}\,\sum_{\langle i,j \rangle}\left(S_i^{+} \, S_j^{-} + S_i^{-} \, S_j^{+} \right)$
since propagation to outside the loop would have to proceed via an empty site of a triangle
where the two other sites are occupied and have alternating sign. These opposite
signs give rise to an exact cancellation, and this destructive interference ensures
that the excitation remains localized on the loop $\ell$
\item The part $J\,\Delta\, \sum_{\langle i,j \rangle} S_i^z \, S_j^z$ is diagonal in the $z$-basis
that we are using here. For a homogeneous system, this term would yield just a constant and would
yield no further conditions. However, for a system with boundaries, the diagonal entries
would depend on the neighborhood and thus yield further conditions.
\end{enumerate}
The first of these two conditions is satisfied for all three loop configurations
sketched in figure~\ref{fig:1loop}. However, only the leftmost example in figure~\ref{fig:1loop}
also satisfies the second condition and thus yields an eigenstate of the spin-$s$ XXZ model.
On the other hand, the second condition is violated by the examples in the middle and on
the right of figure~\ref{fig:1loop} so that they yield eigenstates only for the XY model,
{ i.e.}, $\Delta=0$. Since here we are particularly interested in open boundary
conditions, in what follows we restrict ourselves to the XY model, { i.e.}, $\Delta=0$, 
in order to ensure that any allowed loop configuration yields an exact eigenstate.

\section{Multi-loop configurations}

\label{sec:loop}

Remarkably, the exact localization of the single-magnon states equation~(\ref{eq:1loopState})
on loops facilitates the construction of a macroscopic number of exact multi-magnon
states. Given the configurations of $n_\ell$ loops $\ell_i$, the state 
\begin{equation}
\state{\{\ell_i\}} = \prod_{i=1}^{n_\ell} \left(
\sum_{x_i\in \ell_i} (-1)^{x_i} \, S_{x_i}^-
\right) \, \ferrostate
\label{eq:loopState}
\end{equation}
is also an exact eigenstate with energy $n_\ell\,\omega_0$ provided that each loop
$\ell_i$ respects the condition for a localized single-magnon state and that different loops
are separated by at least one empty site.
At the saturation field, $\omega_0(h_{\rm sat}) = 0$ so that all of the states,
equation~(\ref{eq:loopState}), collapse to zero energy.

For a homogeneous system, { i.e.}, periodic boundary conditions, it has 
been proven that the states, equation~(\ref{eq:loopState}), are not only exact eigenstates, as we have 
explained here, but actually ground states in the respective subspace of 
$S^z$ \cite{SSRS1,Schmidt2002}. Although the situation of open boundary 
conditions that we are considering here does not satisfy the conditions 
of the proof, we find numerically that the states, equation~(\ref{eq:loopState}), are still ground states 
and not just some eigenstates of the XY model on open systems in the 
respective subspace of $S^z$.

Thus, investigation of the ground-state manifold of the model 
equation~(\ref{eq:H}) boils down to studying the geometric problem of 
classical loop configurations $\{\ell_i\}$ on the kagome lattice, at 
least for the subset of ground states given by equation~(\ref{eq:loopState}).

\subsection{Linear relations}

\begin{figure}[tb!]
\centering
\begin{minipage}[c]{0.0cm}
\end{minipage}%
\lower.076\columnwidth\hbox{\includegraphics[height=0.17\columnwidth]{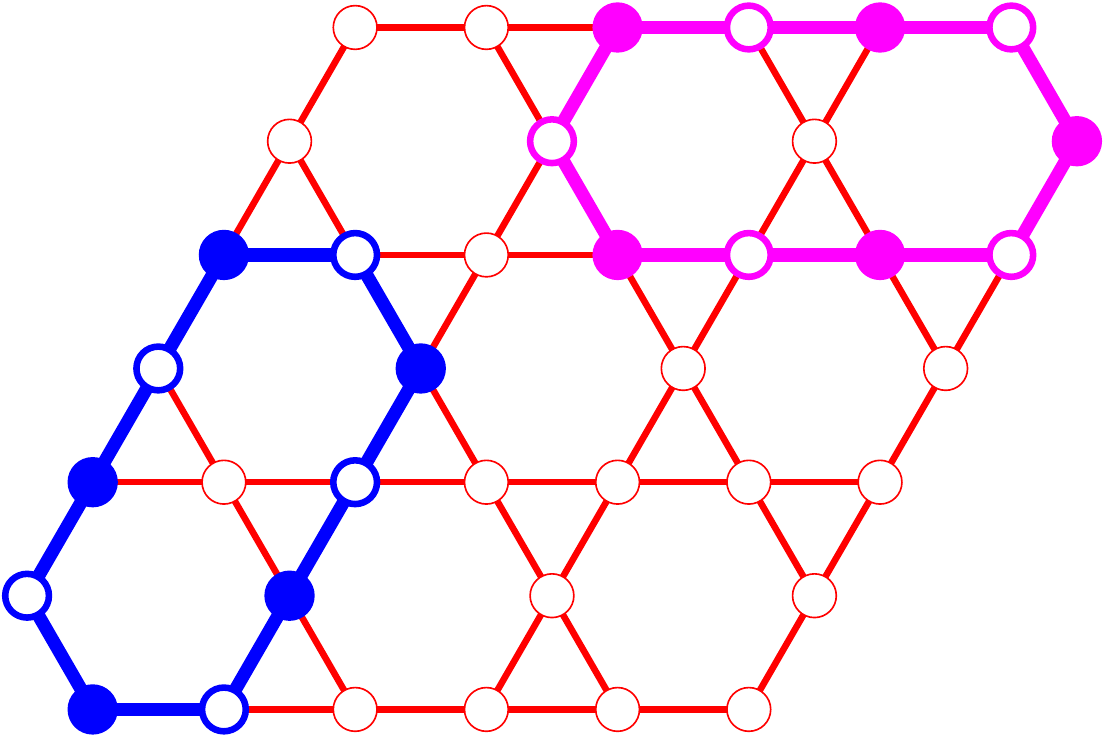}}%
\begin{minipage}[c]{0.5cm}
$=$
\end{minipage}%
\lower.076\columnwidth\hbox{\includegraphics[height=0.17\columnwidth]{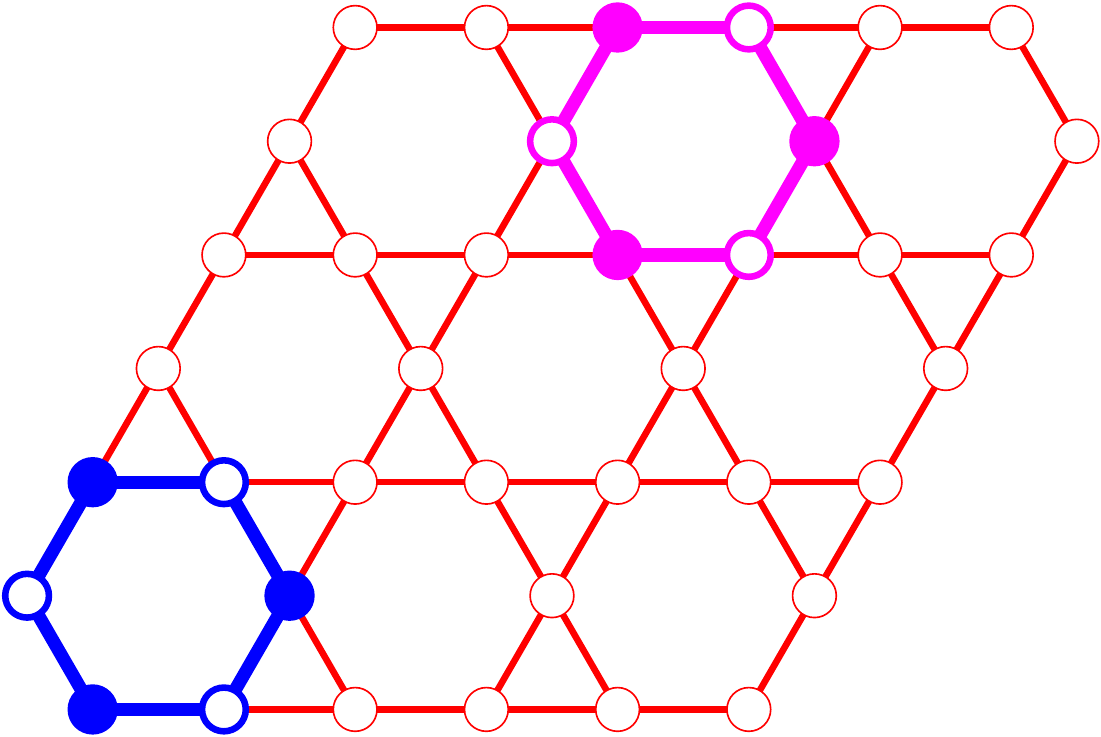}}%
\hspace*{-2mm}
\begin{minipage}[c]{0.5cm}
$+$
\end{minipage}%
\hspace*{-4mm}
\lower.076\columnwidth\hbox{\includegraphics[height=0.17\columnwidth]{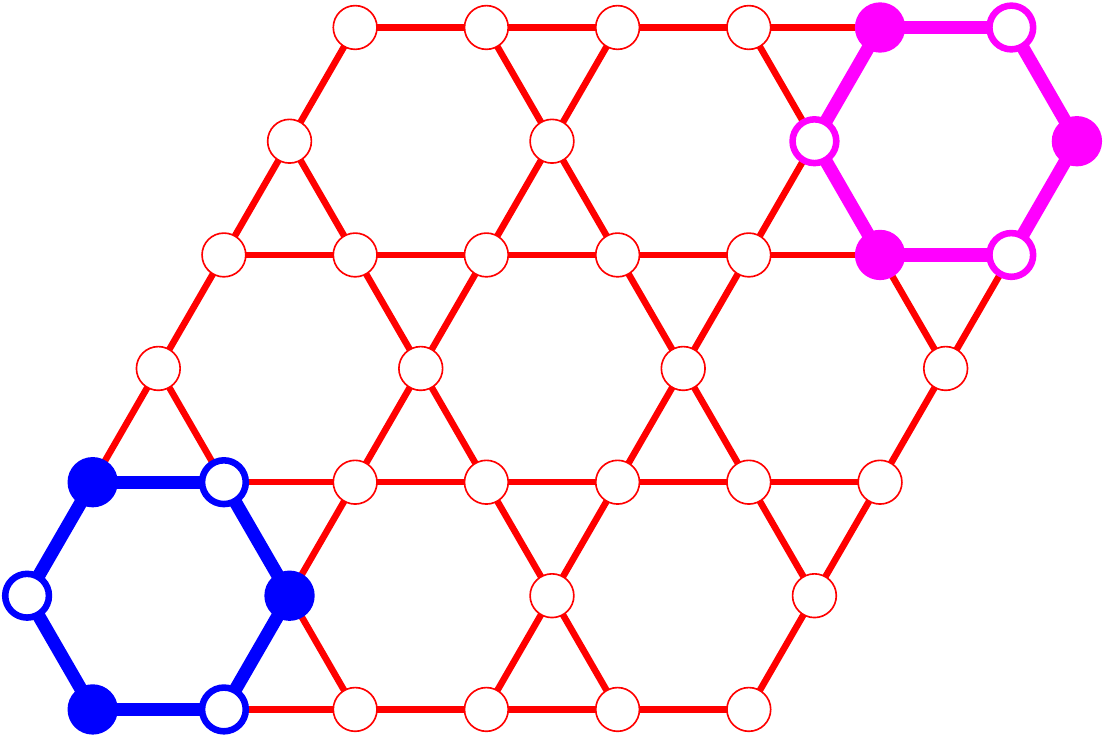}} \\
\hspace*{4.2cm}
\begin{minipage}[c]{0.5cm}
$+$
\end{minipage}%
\hspace*{-5mm}
\lower.076\columnwidth\hbox{\includegraphics[height=0.17\columnwidth]{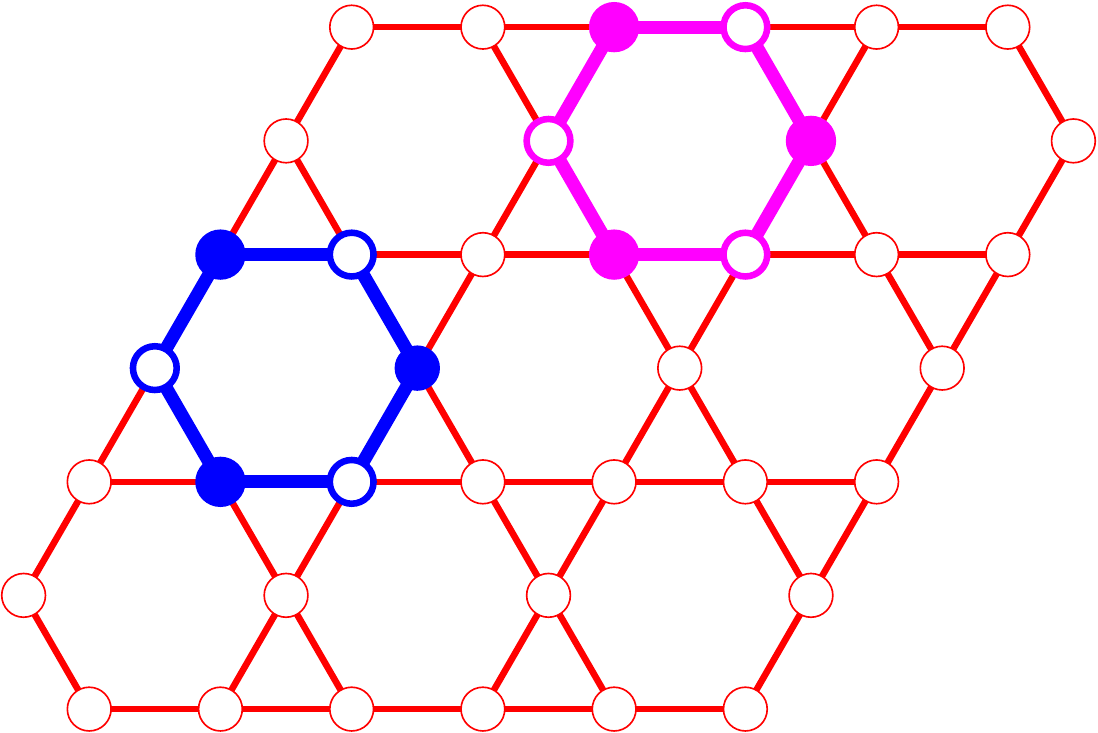}}%
\hspace*{-2mm}
\begin{minipage}[c]{0.5cm}
$+$
\end{minipage}%
\hspace*{-4mm}
\lower.076\columnwidth\hbox{\includegraphics[height=0.17\columnwidth]{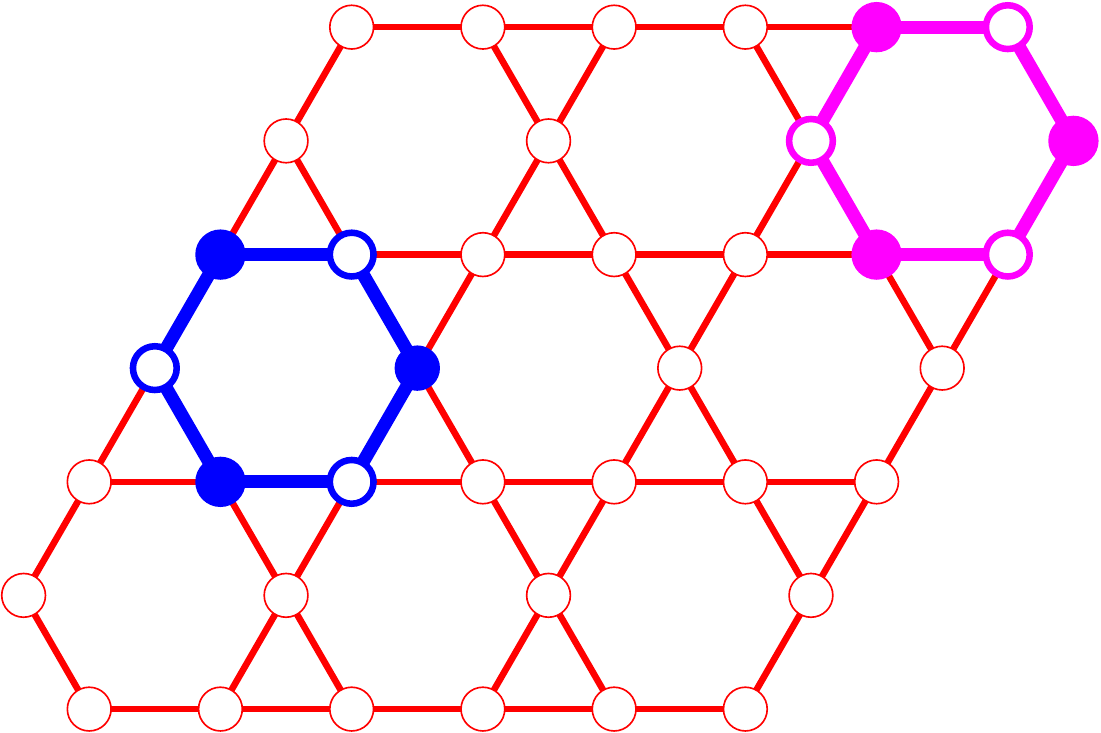}}
\caption{(Colour online) A linear relation between five two-loop configurations on an 
$N=38$ kagome lattice. Sites and bonds belonging to a loop $\ell$ are 
shown in blue and magenta for the first and second loop, respectively 
(bold lines correspond to bonds of a loop). The filled and open circles 
correspond to the alternating signs along the sites of the loops in 
equation~(\ref{eq:loopState}).
\label{fig:2loopRelation}
}
\end{figure}

The states, equation~(\ref{eq:loopState}), are quantum-mechanical wave 
functions so that they may be subjected to linear relations. 
Figure~\ref{fig:2loopRelation} illustrates the relation between two-loop 
configurations. We distinguish here  two different loops by different 
colors, but note that the assignment of first and second loop is 
arbitrary, { i.e.}, that any permutation of the loops in 
equation~(\ref{eq:loopState}) yields the same state. Thanks to relations 
such as the one illustrated in figure~\ref{fig:2loopRelation}, loops in 
any non-nested configurations can be contracted to ``compact'' loops 
residing on a hexagon. In this manner, one recovers the hard-hexagon 
description of all configurations of compact loops 
\cite{ZhT:PRB04,ZhT:PTPS05}. The hard-hexagon states in turn are linearly 
independent, at least in the case of open boundary conditions that we 
consider here\footnote{In the case of periodic boundary conditions, there 
are actually linear relations between the hard-hexagon states on the 
torus, but from the point of view of counting, this deficit is compensated 
by the state with $\omega_-(\vec{0}) = \omega_0$, see also 
figure~\ref{fig:Kenergy} \cite{DRH:LTP07,SSHR20}. }.

\begin{figure}[tb!]
\centering
\includegraphics[height=0.21\columnwidth]{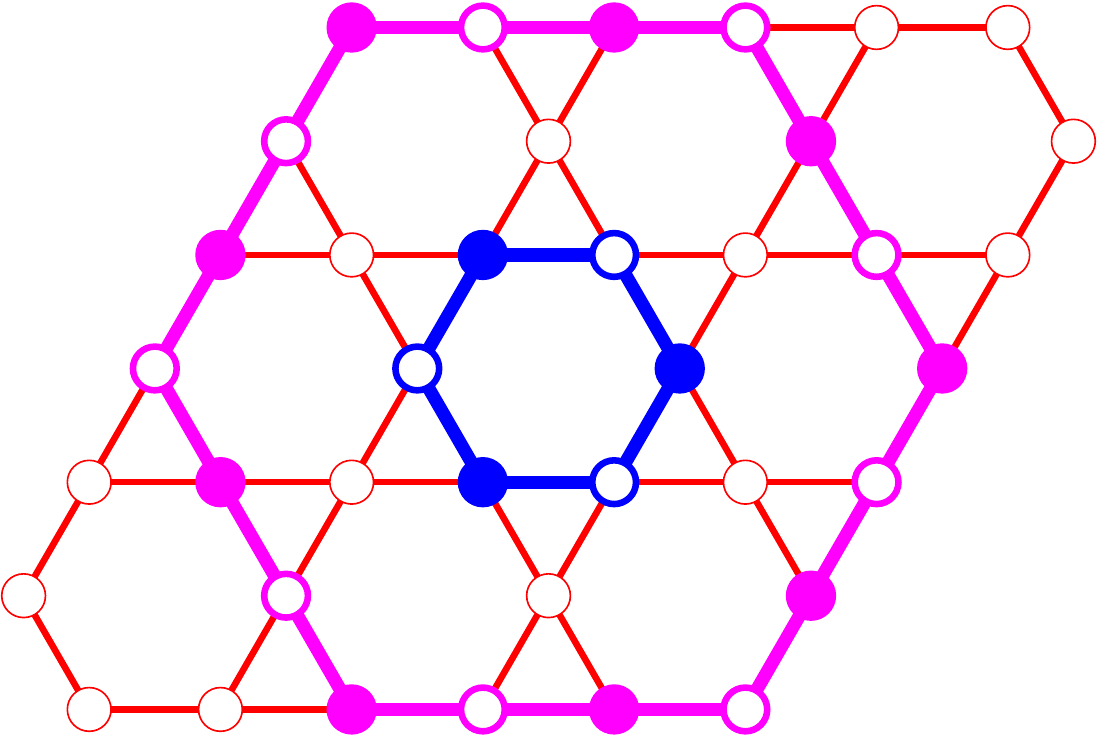}\
\caption{(Colour online) A nested two-loop configuration on an $N=38$ kagome lattice.
Loops are designated in the same way as in figures~\ref{fig:1loop} and \ref{fig:2loopRelation}.
\label{fig:2loopNested}
}
\end{figure}

While hard-hexagon configurations are independent, they do not form a 
basis for the localized-magnon subspace since there exist further nested 
loop configurations that cannot be expressed in terms of hard hexagons 
\cite{ZhT:PRB04,ZhT:PTPS05}. The simplest example for the kagome lattice 
is the nested two-loop configuration shown in 
figure~\ref{fig:2loopNested}. This nesting cannot only be iterated, but 
each new nested loop configuration should be added to an eventual 
classical loop gas description so that we may argue that these nested 
configurations give rise to another macroscopic contribution to the 
ground-state degeneracy \cite{DRH:LTP07}.

We are thus faced with characterizing the loop gas that corresponds to the 
linearly independent states among the equation~(\ref{eq:loopState}). 
Previously, we  observed relations \cite{SSHR20} for which we 
currently have no geometric interpretation. Thus, at this point we rather 
go back to the wave functions equation~(\ref{eq:loopState}) and examine 
the linear relations between them. In the following subsection we  
present the results obtained on a computer for finite open lattices.

\subsection{Counting the loop states}

\label{sec:Counting}

\begin{figure}[htb!]
\centering
\begin{minipage}[t]{1.5cm}
$N_{\rm hex}=4$ \\
$N=19$
\end{minipage}\hspace*{-1.4cm}%
\lower.21\columnwidth\hbox{\includegraphics[height=0.21\columnwidth]{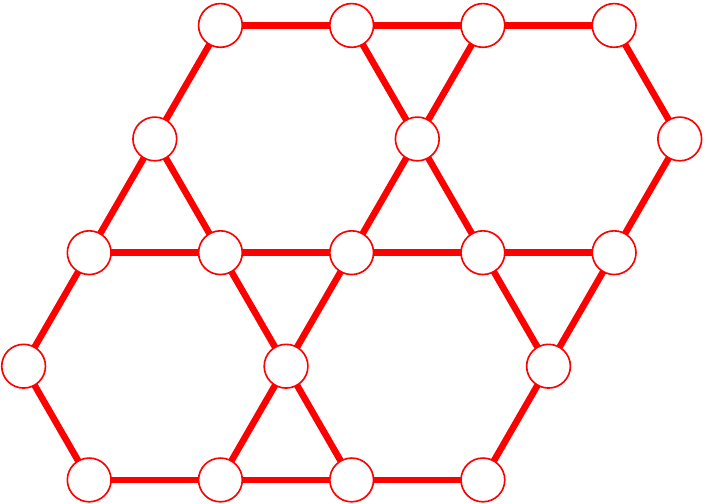}}\quad%
\begin{minipage}[t]{1.5cm}
$N_{\rm hex}=7$ \\
$N=30$
\end{minipage}\hspace*{-1.0cm}%
\lower.21\columnwidth\hbox{\includegraphics[height=0.21\columnwidth]{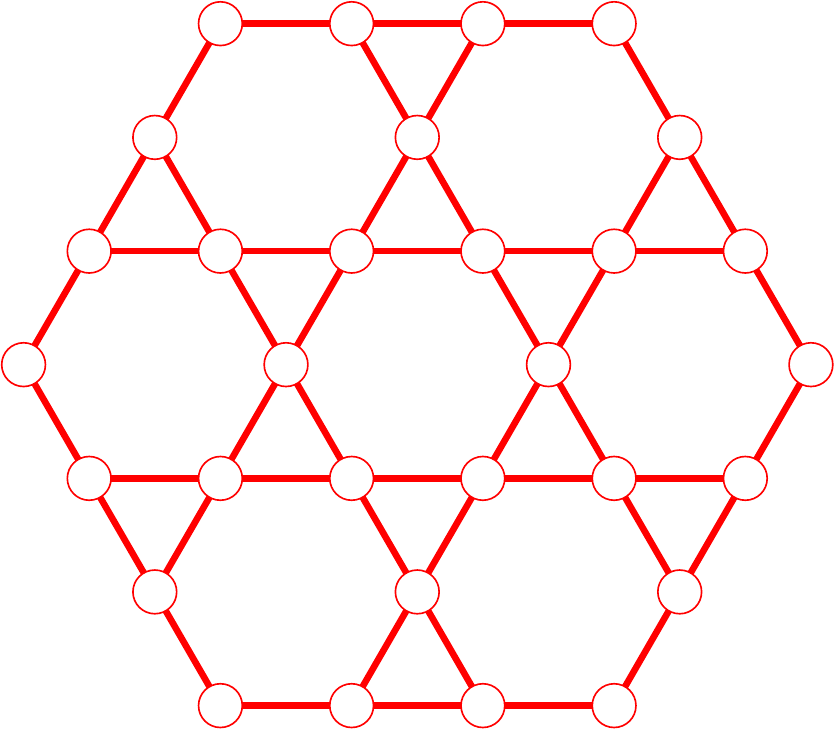}}\quad%
\begin{minipage}[t]{1.5cm}
$N_{\rm hex}=9$ \\
$N=38$
\end{minipage}\hspace*{-1.4cm}%
\lower.21\columnwidth\hbox{\includegraphics[height=0.21\columnwidth]{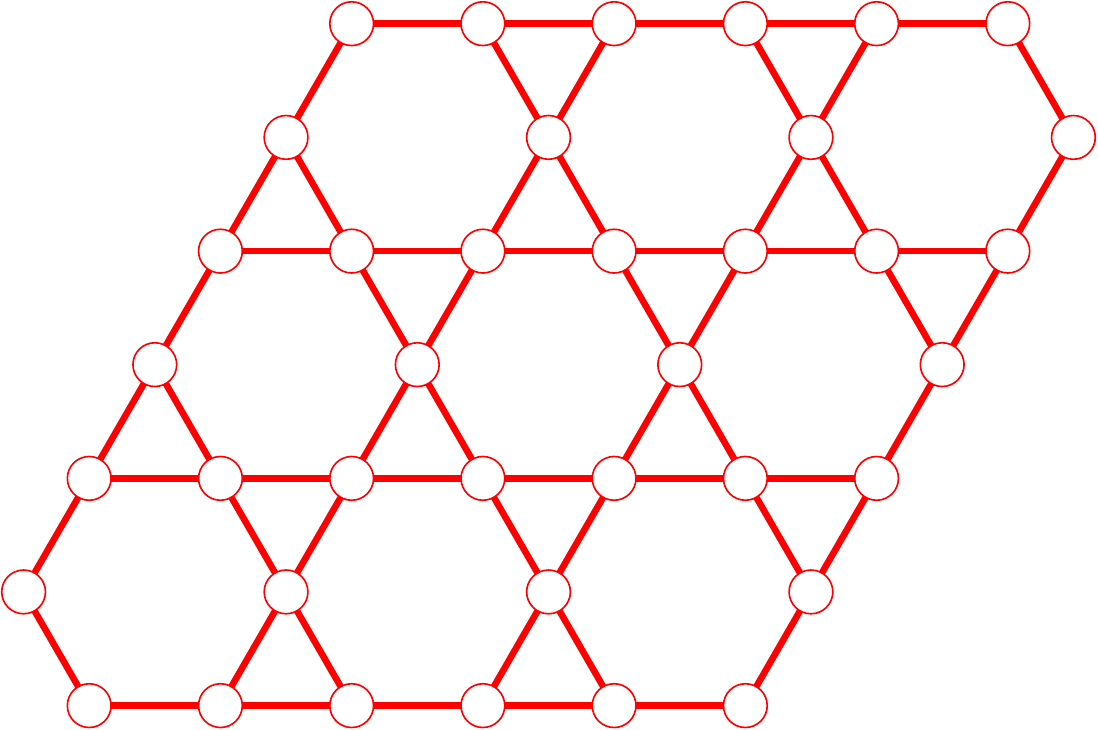}}\\[4mm]
\begin{minipage}[t]{1.5cm}
$N_{\rm hex}=16$ \\
$N=63$
\end{minipage}\hspace*{-1.8cm}%
\lower.3\columnwidth\hbox{\includegraphics[height=0.3\columnwidth]{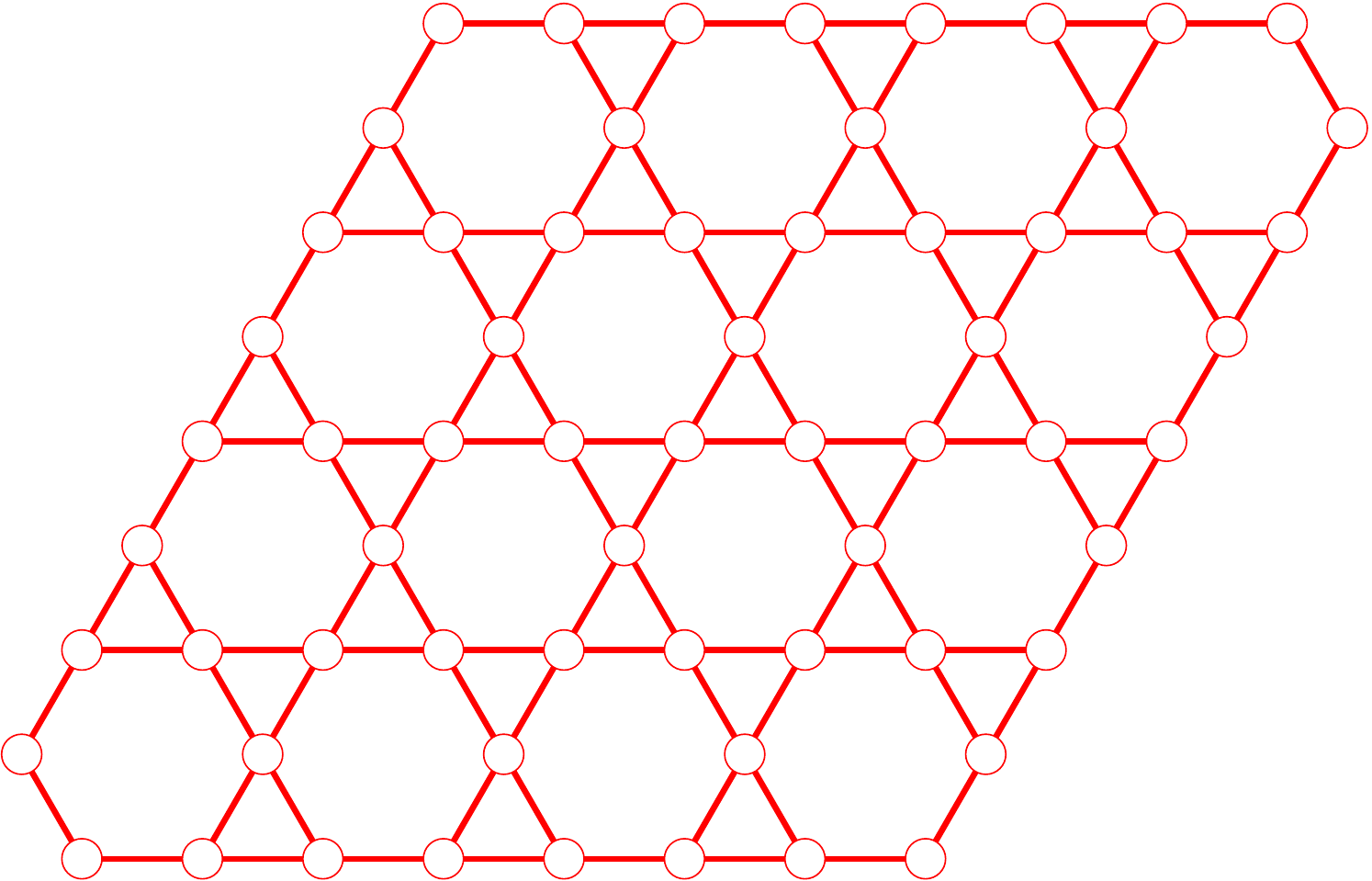}}\qquad%
\begin{minipage}[t]{1.5cm}
$N_{\rm hex}=19$ \\
$N=72$
\end{minipage}\hspace*{-1.1cm}%
\lower.3\columnwidth\hbox{\includegraphics[height=0.3\columnwidth]{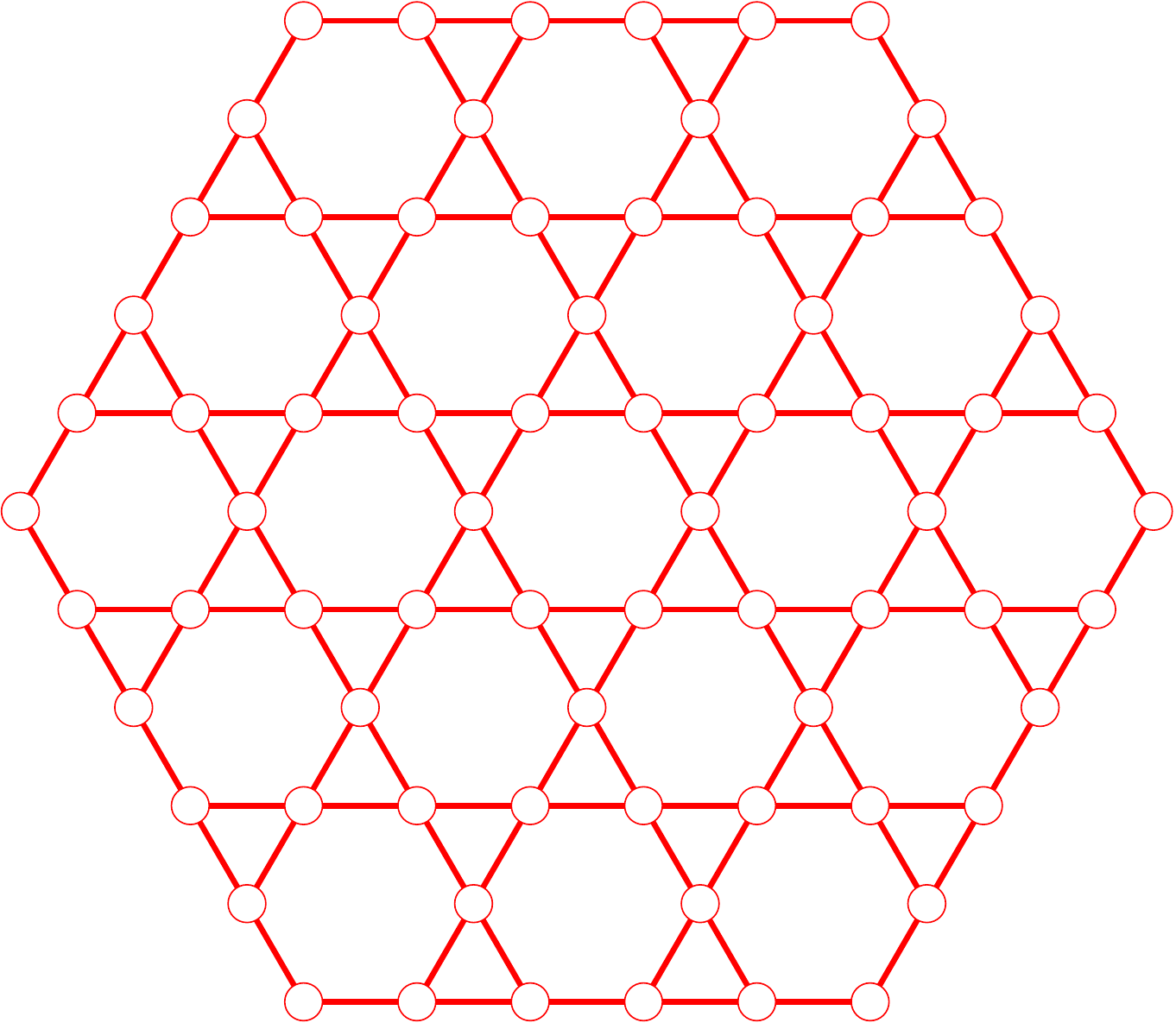}}
\caption{(Colour online) Kagome lattices with open boundary conditions. These include
three
lattices with $N_{\rm hex}=n\times n$ hexagons ($n=2$, $3$, and $4$)
and two hexagonal-shaped lattices. 
\label{fig:lat}
}
\end{figure}

In this section we present the results for the loop gas obtained by 
computer enumeration on the finite kagome lattices with open boundary 
conditions shown in figure~\ref{fig:lat}. These include three cases 
consisting of $N_{\rm hex}=n\times n$ hexagons and $N=3\,n^2+4\,n-1$ 
sites, and two hexagonal-shaped samples containing $N_{\rm hex}=7$ and 
$19$ hexagons ($N=30$ and $72$ spins, respectively). As in previous work 
\cite{SSHR20} for periodic boundary conditions, we started with enumerating 
all allowed $n_\ell$-loop configurations $\{\ell_i\}$. We then make use of 
the fact that, apart from a global sign, the scalar products between the 
states equation~(\ref{eq:loopState}) can be computed just with the 
knowledge of the loop configurations $\{\ell_i\}$ in order to perform a 
Gram-Schmidt orthogonalization of the states equation~(\ref{eq:loopState}) 
and thus eliminate the linearly dependent wave functions.

A key goal of the present work is to compare the resulting dimensions with 
the ground-state degeneracy of the XY model, equation~(\ref{eq:H}) with 
$\Delta = 0$. For this purpose, we take the example of spin 1/2 and 
classify the states according to total $S^z$ and their quantum numbers 
under the point group of the lattices shown in figure~\ref{fig:lat} 
(rotation and/or reflections). As long as the dimension does not much exceed 
$10^5$, we use a library routine to find all the eigenvalues of the 
spin-1/2 XY model in the corresponding sector. In order to push this a bit 
further, we follow the strategy described in reference~\cite{CMP09} to 
obtain a large number of low-lying eigenvectors. We first perform a 
Lanczos procedure \cite{Lanczos:1950} with a relatively large number of 
iterations. The ghosts that are generated by this procedure are projected 
out by re-orthogonalizing the eigenvectors. The main challenge that we 
face here is that we are interested in unusually high ground-state 
degeneracies. In order to ensure that these are found correctly, a 
significant amount of post-processing turns out to be needed (that we 
perform here mainly along the lines of additional vector iteration 
operations) which is the main limiting factor to how far we can push this.


\begin{table}[t!]
\caption{Data for the system consisting of $2\times2 = 4$ hexagons ($N=19$).
For the spin-1/2 XY model, we quote the gap $g$ to the lowest excited
state above the degenerate ground-state manifold in the corresponding
sector of $S^z$.}
\centering
\begin{tabular}{ccc|ccc|c}
\multicolumn{3}{c|}{XY model} &
\multicolumn{3}{c|}{loop gas} &
hard hexagons \\ \hline
$S^z$ & degeneracy & gap $g/J$ & $n_\ell$ & \#\ confs.\
 & \#\ lin.\ indep.
 & \# confs.\ \\
 \hline
$17/2$ & $4$ & $0.14149$ & $1$ & $11$ & $4$ & $4$ \\
$15/2$ & $1$ & $0.02550$ & $2$ & $1$  & $1$ & $1$ \\
\end{tabular}
\label{tab:2x2hex}
\end{table}

\begin{table}[!t]
\caption{Data for the system consisting of 7 hexagons ($N=30$).}\vspace{0.3cm}
\centering
\begin{tabular}{ccc|ccc|c}
\multicolumn{3}{c|}{XY model} &
\multicolumn{3}{c|}{loop gas} &
hard hexagons \\ \hline
$S^z$ & degeneracy & gap $g/J$ & $n_\ell$ & \#\ confs.\
 & \#\ lin.\ indep.
 & \# confs.\ \\
 \hline
$14$ & $7$  & $0.12460$ & $1$ & $88$ & $7$  & $7$ \\
$13$ & $10$ & $0.03021$ & $2$ & $31$ & $10$ & $9$ \\
$12$ & $2$  & $0.02166$ & $3$ & $2$  & $2$  & $2$ \\
\end{tabular}
\label{tab:7hex}
\end{table}

\begin{table}[!t]
\caption{Data for the system consisting of $3\times3 = 9$ hexagons ($N=38$).}
\vspace{0.3cm}
\centering
\begin{tabular}{ccc|ccc|c}
\multicolumn{3}{c|}{XY model} &
\multicolumn{3}{c|}{loop gas} &
hard hexagons \\ \hline
$S^z$ & degeneracy & gap $g/J$ & $n_\ell$ & \#\ confs.\
 & \#\ lin.\  indep.
 & \# confs.\ \\
 \hline
$18$ & $9$  & $0.06853$ & $1$ & $276$ & $9$  & $9$  \\
$17$ & $21$ & $0.01743$ & $2$ & $198$ & $21$ & $20$ \\
$16$ & $11$ & $0.01033$ & $3$ & $32$  & $11$ & $11$ \\
$15$ & $1$  & $0.02243$ & $4$ & $1$   & $1$  & $1$  \\
\end{tabular}
\label{tab:3x3hex}

\end{table}

\begin{table}[!t]
\caption{Data for the system consisting of $4\times4 = 16$ hexagons ($N=63$).}
\vspace{0.3cm}
\centering
\begin{tabular}{ccc|ccc|c}
\multicolumn{3}{c|}{XY model} &
\multicolumn{3}{c|}{loop gas} &
hard hexagons \\ \hline
$S^z$ & degeneracy & gap $g/J$ & $n_\ell$ & \#\ confs.\
 & \#\ lin.\ indep.
 & \# confs.\ \\
 \hline
$61/2$ & $16$  & $0.04087$ & $1$ & $18314$ & $16$  & $16$  \\
$59/2$ & $91$  & $0.01561$ & $2$ & $27966$ & $91$  & $87$  \\
$57/2$ & $207$ & $0.00350$ & $3$ & $15402$ & $207$ & $196$ \\
$55/2$ & $178$ & $0.00364$ & $4$ & $3583$  & $178$ & $176$ \\
$53/2$ &  $46$     & $0.00390$ & $5$ & $212$   & $46$  & $46$  \\
$51/2$ &       &           & $6$ & $2$     & $2$   & $2$   \\
\end{tabular}
\label{tab:4x4hex}
\end{table}

\begin{table}[!t]
\caption{Data for the system consisting of 19 hexagons ($N=72$).}
\vspace{0.3cm}
\centering
\begin{tabular}{ccc|ccc|c}
\multicolumn{3}{c|}{XY model} &
\multicolumn{3}{c|}{loop gas} &
hard hexagons \\ \hline
$S^z$ & degeneracy & gap $g/J$ & $n_\ell$ & \#\ confs.\
 & \#\ lin.\ indep.
 & \# confs.\ \\
 \hline
$35$ & $19$  & $0.05575$ & $1$ & $141203$ & $19$  & $19$  \\
$34$ & $136$ & $0.02159$ & $2$ & $213939$ & $136$ & $129$ \\
$33$ & $430$ & $0.00459$ & $3$ & $127231$ & $430$ & $390$ \\
$32$ & $576$ & $0.00249$ & $4$ & $36986$  & $576$ & $532$ \\
$31$ &       &           & $5$ & $4518$   & $303$ & $297$ \\
$30$ &       &           & $6$ & $210$    & $55$  & $55$  \\
$29$ &       &           & $7$ & $2$      & $2$   & $2$   \\
\end{tabular}
\label{tab:19hex}
\end{table}

Results for the lattices of figure~\ref{fig:lat} are summarized in tables 
\ref{tab:2x2hex}--\ref{tab:19hex}. We note that some empty entries for the 
spin-1/2 XY model in tables \ref{tab:4x4hex} and \ref{tab:19hex} were 
beyond our numerical capacities. Before we discuss the ground-state 
degeneracies, we mention that these tables also quote a value of the gap 
$g$ between the ground-state manifold and the first excited state in the 
corresponding sector of $S^z$ for the spin-1/2 XY model. One observes that 
this gap can become quite small so that one may expect excited states to 
become relevant for the finite-temperature properties that we discuss 
in section \ref{sec:finiteT}.

In the present section we focus on the ground-state properties encoded in 
tables \ref{tab:2x2hex}--\ref{tab:19hex}. First we observe that a maximum 
of $n_\ell \leqslant (N_{\rm hex}+2)/3$ loops can be packed on the lattices of 
figure~\ref{fig:lat}. The maximum packing corresponds to a ``magnon 
crystal'' where the localized excitations reside on hexagons that are 
separated from each other by empty sites. For the open systems that we 
consider here, we can firstly pack sightly more than $N_{\rm hex}/3$ 
loops into the system by pushing the loops to the boundaries. Secondly, the 
degeneracy in this maximum sector of $n_\ell$ is not 3 as for 
periodic boundary conditions \cite{SSHR20}, but rather reduced to 1 or 2, 
depending on whether this densest packing is symmetric under the 
point-group symmetry or whether there are two symmetry-related closest 
packings.

Returning to smaller numbers of loops, we observe that the maximum number 
of geometrically allowed loop configurations before considering the linear 
relations is in the sectors with $n_\ell = 1$ or $2$ for the present 
cases. Given that these numbers are large, these enumerations constitute a 
relevant limiting factor for our enumeration procedure, but they are 
necessary as a basis of the enumeration in the sectors with a higher 
number of loops.

If we now consider the linearly independent loop states, the most 
important observation probably is that their number is exactly equal to 
the number of ground states of the spin-1/2 XY model in all cases where we 
are able to make such a comparison. This suggests that the additional 
states that appeared for periodic boundary conditions \cite{SSHR20} 
disappear when one goes to open ones so that for open boundary 
conditions the states, equation~(\ref{eq:loopState}), do indeed span the 
ground-state manifold of the model equation~(\ref{eq:H}) in its high-field 
regime. In view of this equality, we  focus on a comparison of the 
linearly independent loop states and hard hexagons.

The smallest nested loop configuration is the one shown in 
figure~\ref{fig:2loopNested}. It occupies 30 sites (plus possibly 
surrounding empty sites that are required to separate it from neighboring 
loops). Since even the smallest nested configuration does not fit on the 
$N=19$ lattice of figure~\ref{fig:lat}, all degeneracies are equal in this 
case, as one observes in table \ref{tab:2x2hex}. The configuration of 
figure~\ref{fig:2loopNested} fits exactly once on the $N=30$ and $N=38$ 
lattices of figure~\ref{fig:lat} such that one obtains exactly one further 
linearly independent loop state for these two lattices in the sector with 
$n_\ell = 2$, see tables \ref{tab:7hex} and \ref{tab:3x3hex}. There are 
more non-hard-hexagon states for the $N=63$ and $72$ lattices of 
figure~\ref{fig:lat} although the differences that we observe in tables 
\ref{tab:4x4hex} and \ref{tab:19hex} remain relatively small. Indeed, 
bigger lattices would be needed in order to allow more multi-magnon 
configurations built on nested loops when we impose open boundary 
conditions, as also the following consideration shows. Compact loops 
occupy  6 sites around one hexagon while the smallest loop that 
encloses such a compact loop requires 18 sites, compare again 
figure~\ref{fig:2loopNested}, and to construct a third nested loop, one 
needs to put it on 30 boundary sites of the $N=72$ lattice in 
figure~\ref{fig:lat}. The preceding considerations also explain why the 
difference between linearly independent loop states and hard-hexagon 
configurations is maximal for intermediate loop densities, since these 
leave more space for bigger loops.

\begin{figure}[tb!]
\centering
\includegraphics[width=9 true cm]{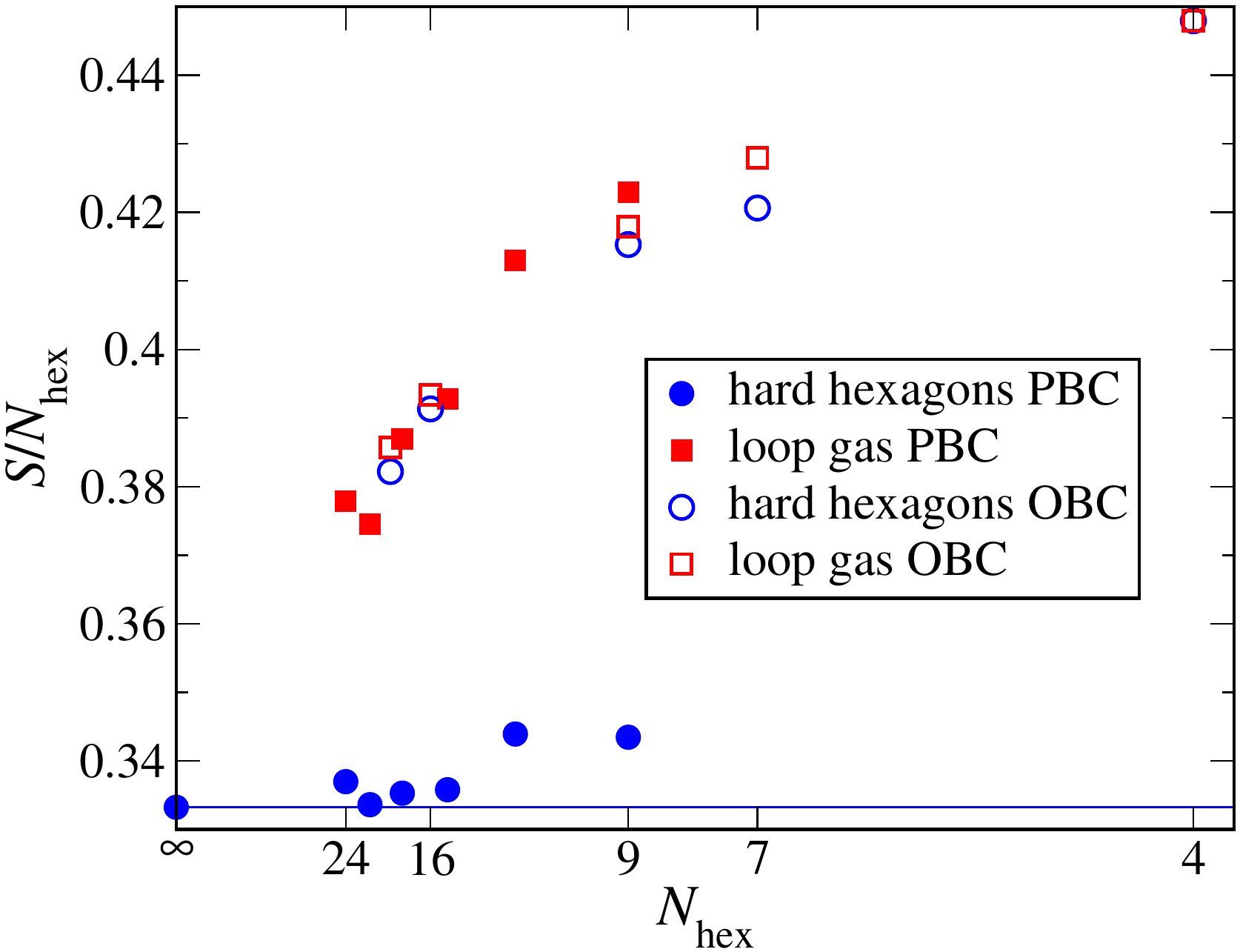}
\caption{(Colour online) Total entropy $S/N_{\rm hex}$ of ground states per hexagon.
We show both the results for open boundary conditions (OBC) corresponding to
tables \ref{tab:2x2hex}--\ref{tab:19hex} as well as previous
results for periodic boundary conditions (PBC) \cite{SSHR20}.
The horizontal line denotes the known thermodynamic limit for
hard hexagons \cite{BaxterTsang1980}.
\label{fig:S}
}
\end{figure}

The results of tables \ref{tab:2x2hex}--\ref{tab:19hex} are summarized
in figure\ \ref{fig:S} that shows the total entropy corresponding to the
linearly independent loop configurations and hard hexagons.
We also include previous results for periodic boundary conditions
\cite{SSHR20} for comparison. Here, we choose a normalization per
hexagon which in the case of open boundary conditions is more
appropriate than normalizing per site while for periodic
boundary conditions these normalizations are equivalent since
in that case $N_{\rm hex} = N/3$. The horizontal line in figure~\ref{fig:S}
shows the known result for the thermodynamic limit of hard hexagons
\cite{BaxterTsang1980}
\begin{equation}
\lim_{N_{\rm hex} \to \infty} \frac{S_{\rm hard\ hexagons}}{N_{\rm hex}}
 \approx  0.333 242 721 976 \, .
\label{eq:BaxterShh}
\end{equation}
Comparison with  the results for finite $N_{\rm hex}$ shows that for 
hard hexagons open boundary conditions clearly gives rise to bigger 
finite-size effects than periodic ones. For the loop gas, boundary 
conditions seem to matter less, but conclusions about the thermodynamic 
limit are evidently difficult to draw based on the data for finite $N_{\rm 
hex}$. We observe again that for open boundary conditions and the system 
sizes that we have studied here, the entropy of the loop gas is only 
slightly higher than that of hard hexagons. However, we recall that when 
including nested loop configurations \cite{ZhT:PRB04,ZhT:PTPS05} into the 
many-particle physics, we expect another macroscopic contribution to the 
number of ground-state configurations \cite{DRH:LTP07}. Consequently, the 
results for $S/N_{\rm hex}$ of the loop gas and hard hexagons should 
separate more clearly for systems that are bigger than those discussed 
here, so that the entropy for the loop gas converges to a value that is bigger than 
equation~(\ref{eq:BaxterShh}) in the thermodynamic limit.

We comment furthermore that $S/N_{\rm hex}$ corresponds to the 
ground-state entropy of the spin-1/2 XY model exactly at the saturation 
field equation~(\ref{eq:Hsat}) since $n_\ell\,\omega_0(h_{\rm sat}) = 0$ 
independently of the number of loops $n_\ell$. Since 
equation~(\ref{eq:BaxterShh}) is a lower bound for the thermodynamic limit 
of this entropy, we conclude that the thermodynamic limit of the entropy 
$\lim_{N_{\rm hex}\to\infty} S/N_{\rm hex}$ is non-zero exactly at the 
saturation field so that the third law of thermodynamics is violated in 
this quantum system at $h=h_{\rm sat}$.

Although open boundary conditions do enhance finite-size effects and thus 
complicate the analysis, there is nevertheless an important message to be 
retained from the present analysis, namely that the loop gas seems to 
provide an exact match of the ground-state degeneracy of the spin-1/2 XY 
model. Since boundary conditions should become irrelevant in the 
thermodynamic limit, this implies in turn that the loop gas should also amount 
 to an exact description of the ground-state manifold of the spin-$s$ 
XXZ model with periodic boundary conditions once the thermodynamic limit 
is taken.

\subsection{Contribution of localized states to the thermodynamics}

With the degeneracies of the ground states given in tables 
\ref{tab:2x2hex}--\ref{tab:19hex}, it is straightforward to compute the 
contribution of the hard hexagons, or more generally the loop gas, to 
thermodynamic properties. The corresponding partition function is given by
\begin{equation}
Z_{\rm loc} = \sum_{n_{\ell}=0}^{n_{\max}} {\rm deg}(n_\ell) \, {\rm e}^{-\beta\,n_\ell\,\omega_0}
= \sum_{n_{\ell}=0}^{n_{\max}} {\rm deg}(n_\ell) \, {\rm e}^{-n_\ell\,\beta\,(h-h_{\rm sat})} \, ,
\label{eq:Zloc}
\end{equation}
where $\beta=1/T$ is the inverse temperature (we set $k_{\text{B}}=1$). ${\rm 
deg}(n_\ell)$ is the number of hard-hexagon or linearly independent loop 
configurations, respectively. We comment that the data in tables 
\ref{tab:2x2hex}--\ref{tab:19hex} needs to be complemented by the empty 
system corresponding to the ferromagnetically polarized state, { i.e.}, 
${\rm deg}(0)=1$. The energy of an $n_\ell$-particle state 
$\,n_\ell\,\omega_0$ is rewritten by substituting the saturation 
field according to equation~(\ref{eq:Hsat}) into 
equation~(\ref{eq:singleMagnon}). The result equation~(\ref{eq:Zloc}) 
makes it obvious that all thermodynamic quantities in the hard-hexagon and 
the loop-gas description depend only on the combination $\beta\,(h-h_{\rm 
sat}) = (h-h_{\rm sat})/T$ and not on temperature $T$ and magnetic field 
$h$ separately. The physical reason is that all configurations in a given 
sector of $n_\ell$ have the same energy so that the only characteristic 
energy scales are $T$ and $h-h_{\rm sat}$, and the partition function must 
thus depend on the ratio of these two quantities.

\begin{figure}[tb!]
\centering
\includegraphics[width=9 true cm]{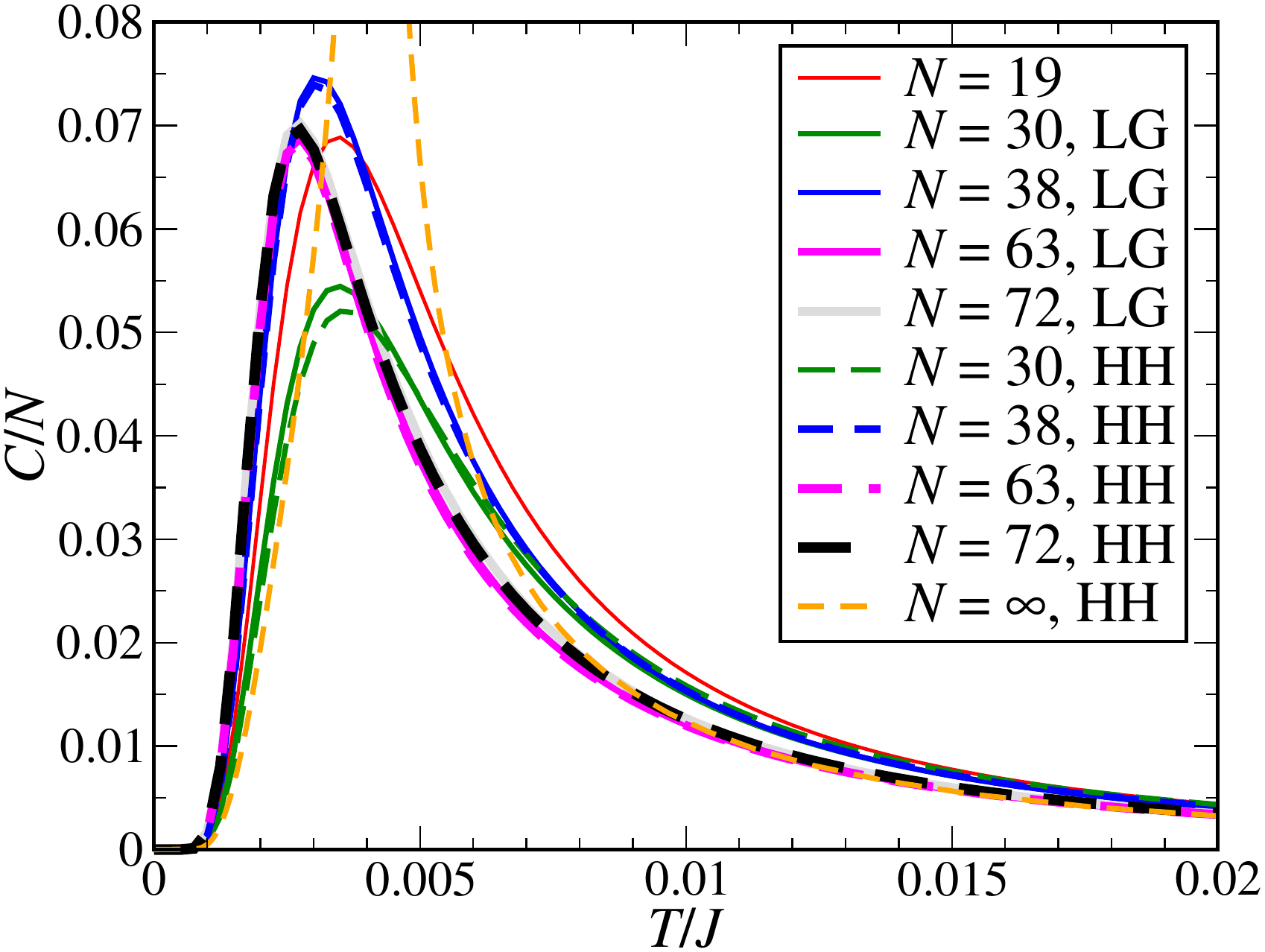}
\caption{(Colour online) Specific heat per site $C/N$ at a magnetic field $h=0.99\,J$ as 
obtained from hard hexagons (HH) and the loop gas (LG). For $N=19$, both 
descriptions yield identical results owing to identical degeneracies, see 
table~\ref{tab:2x2hex}. For hard hexagons, we include the known result for 
the thermodynamic limit $N=\infty$ \cite{ZhT:PRB04,Baxter1980,ZhT:PTPS05}.
\label{fig:ChhLoop}
}
\end{figure}

Figure \ref{fig:ChhLoop} shows the result for the specific heat per site 
$C/N$ in both the hard-hexagon (HH) and loop-gas (LG) picture. For the later 
comparison with numerical data for the spin-1/2 XY model we take the 
magnetic field to be $h=0.99\,J$, but as we have just pointed out, one 
could rescale this to any value $h < h_{\rm sat}$ by rescaling the 
temperature axis accordingly. In the case of hard hexagons, the result for 
the thermodynamic limit $N=\infty$ is known exactly
\cite{ZhT:PRB04,Baxter1980,ZhT:PTPS05}, and we include this 
in figure~\ref{fig:ChhLoop}. This latter curve diverges at the temperature 
$T_c/J \approx 0.004156$ which corresponds to the phase transition from 
the magnon crystal at low temperatures to a paramagnetic disordered phase 
at high temperatures in the hard-hexagon description. The curves for 
finite $N$ exhibit a maximum at a temperature close to the phase 
transition in the infinite system.

A noteworthy observation in figure~\ref{fig:ChhLoop} is that the results 
for hard hexagons and the loop gas are very close for the lattices with 
open boundary conditions that we consider here; for the $N=19$, system they 
are even identical. The reason for this is that the degeneracies are 
either close or even identical, see tables~\ref{tab:2x2hex}--\ref{tab:19hex} and figure~\ref{fig:S}. In particular, 
we do not observe a significant reduction of temperature at the 
maximum of $C$ by additional loop-gas states that we have observed for 
periodic boundary conditions \cite{SSHR20}. Indeed, the system sizes that 
we are able to access here with open boundary conditions probably remain 
too small to accommodate a thermodynamically relevant number of loop 
configurations beyond hard hexagons, as we have already explained in the 
discussion of figure~\ref{fig:S}, { i.e.}, the open boundary conditions 
enhance once again the finite-size effects and one would have to go to 
bigger systems in order to see more clearly the difference between hard hexagons and 
the loop gas.

\section{Spin-1/2 XY model}

In this final chapter, we  compare the contribution of the
localized-magnon states to the low-temperature thermodynamics in the
vicinity of the saturation field with numerical data for the full
spin-1/2 XY model.

Full exact diagonalization is only possible for the smallest lattice of 
figure~\ref{fig:lat}, namely the $N=19$ one. Hence, we resort to other 
methods to compute the thermodynamic properties for bigger lattices. 
The key idea is to perform a Monte-Carlo-like sampling of the Hilbert 
space that we briefly sketch  here using the language of thermal pure 
quantum (TPQ) states~\cite{Sugiura2012,Sugiura2013}. 
One starts at infinite
temperature where  the expectation value of an operator $\mathcal{O}$ is
given by its trace,  that in turn can be approximated by the random trace
average
\begin{equation}
  \label{eq:generictraceaverage}
  {\rm Tr}(\mathcal{O}) = d \, \overline{\astate{r}  \mathcal{O} \state{r}} \, ,
\end{equation}
where $\state{r}$ is a random vector with normal-distributed components, 
$d$ is the dimension of the Hilbert space, and $\overline{\cdots}$ 
denotes averaging over random realizations of $\state{r}$. Thermal 
expectation values are then obtained by an imaginary time evolution
\begin{equation}
  \label{eq:thermalexpectation}
  \langle \mathcal{O} \rangle
  = \frac{\overline{\astate{r} \text{e}^{-\beta\, H / 2} \,
                   \mathcal{O} \, \text{e}^{-\beta\, H / 2} \state{r}}}%
    {\overline{\astate{r} \text{e}^{-\beta H}  \state{r}}} \, ,
\end{equation}
where, as in equation~(\ref{eq:Zloc}), $\beta=1/T$ is the inverse 
temperature and we have set $k_{\text{B}}=1$. The average in 
equation~(\ref{eq:generictraceaverage}) is approximated by a finite and 
small number of $R$ different random realizations of $\state{r}$ in each 
symmetry subspace. As shown in 
references~\cite{Hams2000,Goldstein2006,Sugiura2012,Sugiura2013}, the 
statistical error caused by the random sampling of the vectors $\state{r}$ 
is related to the density of states of the Hamiltonian $H$ and can become 
small when the dimension $d$ of the subspace is large and thus 
exponentially small in the system size $N$.

Certain efficient evaluations of equation~(\ref{eq:generictraceaverage}) 
make use of the Lanczos algorithm~\cite{Lanczos:1950}. A first 
implementation that we have employed \cite{Wietek2019} used the above 
language of TPQ and a Krylov approximation to the imaginary-time 
evolution~\cite{Hochbruck1997}. For details of the parallelization of this 
implementation we refer to~\cite{Wietek2018}.

A different implementation of the above procedure is known under the name 
of finite-temperature Lanczos method (FTLM) 
\cite{JaP:PRB94,JaP:AP00,ScW:EPJB10,PrB:SSSSS13,HaS:EPJB14,ScT:PR17,PRE:COR17,kago42,Accuracy_FTL_PRR2020,Seki_FTLM2020}. 
At the level of the method and with the details that we have provided 
here, TPQ and FTLM are indistinguishable. However, the second 
implementation is quite different from the fist one and it is based on 
J.~Schulenburg's {\it spinpack} code \cite{spin:259,RSHS04}. We 
continue to refer to the first implementation by ``TPQ'' and to the 
second one by ``FTLM''.

\subsection{Zero-temperature magnetization curve}

\begin{figure}[tb!]
\centering
\includegraphics[width=9 true cm]{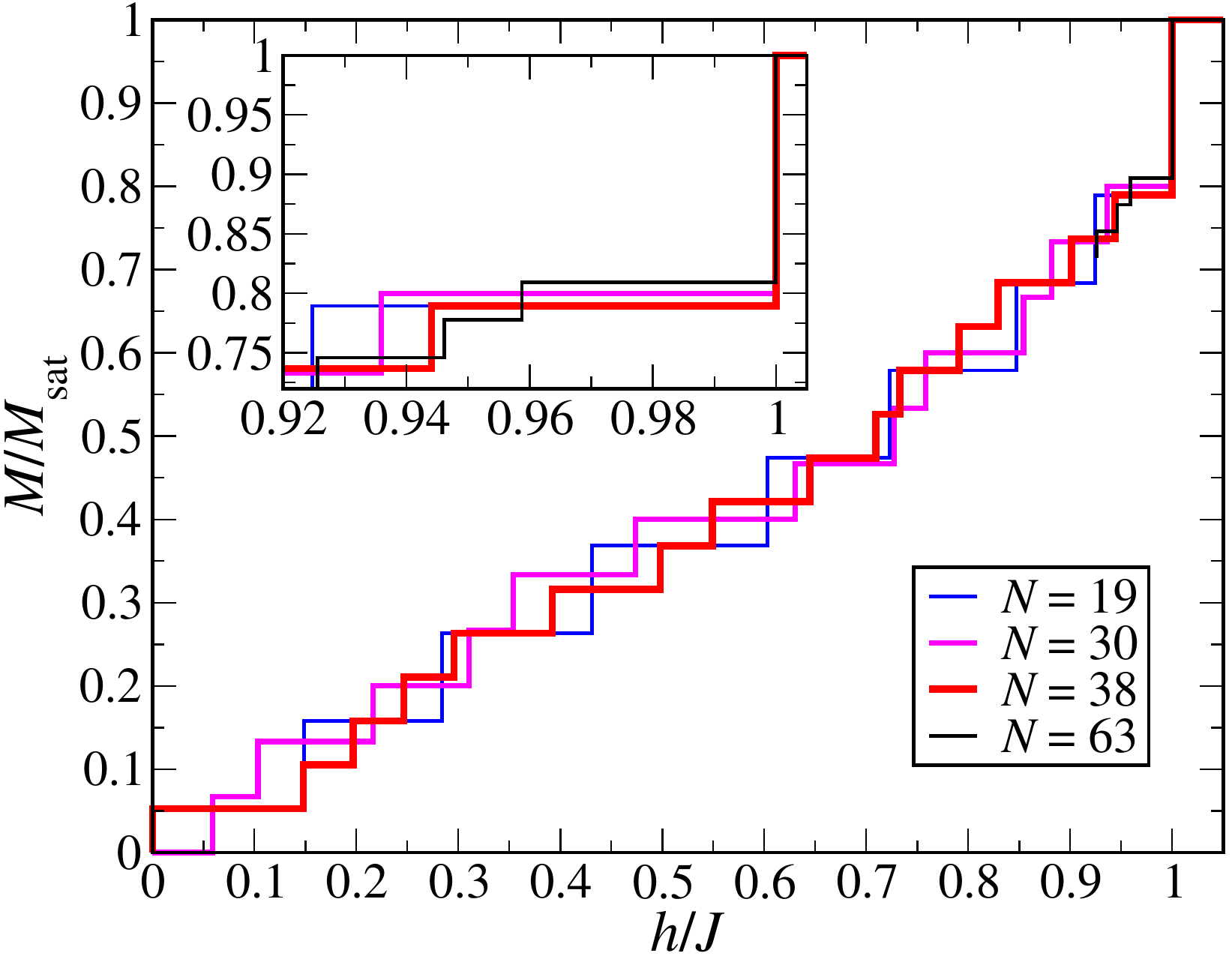}
\caption{(Colour online) Zero-temperature magnetization curve of the spin-1/2
XY model. The magnetization $M$ is normalized to its saturation
value $M_{\rm sat}$. The inset presents a closer view of the
behavior close to saturation.
\label{fig:M}
}
\end{figure}

As a first byproduct, we obtain the ground-state energies and from these 
it is straightforward to reconstruct the zero-temperature magnetization 
curve. Since it provides useful information about the field range where 
localized magnons are low-energy excitations, we start by presenting 
numerical results for the zero-temperature magnetization curve in 
figure~\ref{fig:M} where we normalize the saturation value of the 
magnetization to $1$.

We note that the saturation field in figure~\ref{fig:M} is indeed $h_{\rm 
sat}/J=1$, as expected according to equation~(\ref{eq:Hsat}). The 
transition to saturation occurs via a jump in the magnetization curve. 
Usually, such jumps are associated with a concave piece in the energy $E$ 
as a function of $S^z$ curve $E(S^z)$. The present case is exceptional in 
so far that it is associated not only with a linear piece in $E(S^z)$, but 
that in addition, the ground states in each subspace of $S^z$ are also 
degenerate, as we have discussed in chapters \ref{sec:modelLoc} and 
\ref{sec:loop}, so that the ground-state manifold becomes 
macroscopically degenerate and the third law of thermodynamics is violated 
by this quantum system at $h=h_{\rm sat}$.

Just below this jump in the magnetization curve, we observe a plateau in 
figure~\ref{fig:M}. In the thermodynamic limit, the plateau should have 
the value $M/M_{\rm sat} = 7/9$, as one does indeed find on suitable 
lattices with periodic boundary conditions, see 
references~\cite{SHS:PRL02,NSH:NC13,CDH:PRB13,SSHR20} for the Heisenberg 
model. The deviations from $M/M_{\rm sat} = 7/9 = 0.777\ldots$ that we 
observe in figure~\ref{fig:M} are finite-size effects that arise due to 
the open boundary conditions that we employ here. The wave function of 
this plateau is the aforementioned magnon crystal that is stable in the 
range of magnetic fields where a plateau is observed. By inspection of 
figure~\ref{fig:M} we conclude that the plateau width is approximately 
4\%\ of the saturation field $h_{\rm sat}$, { i.e.}, the 
\emph{relative} stability range of the magnon crystal in the spin-1/2 XY 
model is comparable to that in the spin-1/2 Heisenberg model 
\cite{SHS:PRL02,NSH:NC13,CDH:PRB13,SSHR20}.

Apart from the vicinity of the saturation field, no other distinct 
features are visible in figure~\ref{fig:M}. This applies in particular to 
a possible $M/M_{\rm sat} = 1/3$ plateau that was suggested to persist in 
the XY limit that we are considering here \cite{CGH:PRB05}. However, if 
this $1/3$ plateau exists in the XY limit, it would be expected to be 
narrow \cite{CGH:PRB05} so that the comparably large finite-size effects 
caused by the open boundary conditions can easily obscure it.

\subsection{Finite-temperature properties of the magnon crystal}

\label{sec:finiteT}

Since we concluded from figure~\ref{fig:M} that the magnon crystal state 
is stable at $T=0$ for $0.96\,J \lesssim h \leqslant J$, we select $h=0.99\,J$ 
for a discussion of thermodynamic properties.

\begin{figure}[tb!]
\centering
\includegraphics[width=11 true cm]{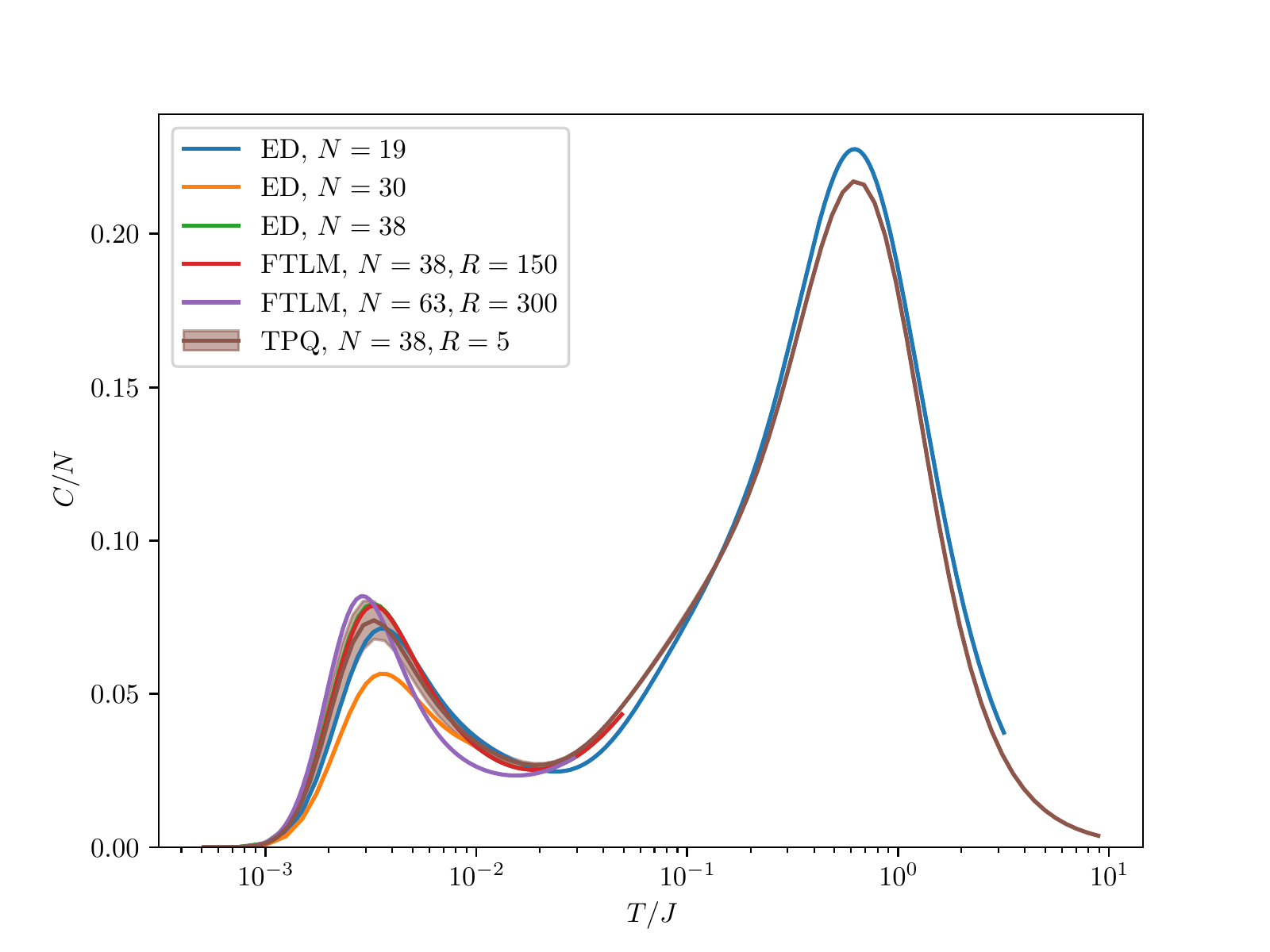}
\caption{(Colour online) Specific heat per site $C/N$ of the spin-1/2 XY model at a 
magnetic field $h=0.99\,J$. Note the logarithmic temperature axis. Here, we 
compare several numerical methods for different system sizes: for $N=19$, 
we have full exact diagonalization (ED), for $N=30$ full diagonalization 
in the sectors with $S^z \geqslant 9$ supplemented by the low-energy part of the 
spectrum in the sectors with $6 \leqslant S^z \leqslant 8$, and for $N=38$ we used 
full exact diagonalization results for $S^z \geqslant 14$ supplemented by the 
low-energy part of the spectrum for $S^z = 13$. These are compared with 
FTLM for $N=38$ and $63$, and TPQ for $N=38$, respectively.
\label{fig:specheat_edsizes}
}
\end{figure}

Figure \ref{fig:specheat_edsizes} presents numerical results for the 
specific heat of the spin-1/2 XY model. In this figure, we compare in 
particular the different methods discussed at the beginning of this 
chapter. $N=19$ is the only system where we can perform full ED and thus 
obtain numerically ``exact'' results over the whole temperature range. For 
$N=30$ and a field close to the saturation field, we can still perform ED 
for the low-energy part of the spectrum since in this parameter regime 
only the sectors with high $S^z$ are relevant and the corresponding 
dimensions are relatively small. For the $N=30$ case shown in 
figure~\ref{fig:specheat_edsizes}, we have completely diagonalized the 
sectors with $S^z\geqslant 9$ and further obtained the low-energy part of the 
spectrum in the sectors with $6 \leqslant S^z \leqslant 8$ using the method outlined 
at the beginning of section \ref{sec:Counting}. This can be pushed till 
$N=38$ where we performed a complete diagonalization of the sectors 
$S^z \geqslant 14$ and determined the low-energy part of the spectrum for $S^z=13$
which includes two sectors below the magnon crystal, see 
table \ref{tab:3x3hex}. For $N=38$, we further performed a TPQ 
computation that takes all sectors of $S^z$ into account and a FTLM 
computation that again focuses on sectors with higher $S^z \geqslant 11$. The 
latter two methods are now subjected to statistical errors that we show for 
TPQ by an error tube in figure~\ref{fig:specheat_edsizes}. Comparison of 
all three data sets for $N=38$ in the temperature regime $T\lesssim 
10^{-2}\,J$ shows that they agree within statistical errors so that we 
can trust each of them. Finally, for $N=63$, 
figure~\ref{fig:specheat_edsizes} includes just an FTLM data set. This 
large system size is again possible because we restrict ourselves to high sectors of 
$S^z \geqslant 47/2$, but this implies that we only have access to the 
low-temperature regime.

Turning now to the physics behind figure~\ref{fig:specheat_edsizes}, first, 
one observes a classical fluctuation peak at a temperature $T$ of the 
order of $J$. The fact that the $N=19$ ED and $N=38$ TPQ curves are close 
at this elevated temperature shows that correlations are still short-range 
so that finite-size effects are small. The most interesting feature in 
the present context is the second peak in the temperature regime $T < 
10^{-2}\,J$ of figure~\ref{fig:specheat_edsizes}. Here, finite-size effects 
are more important. We now focus on this low-temperature peak in the 
specific heat and explain its relation to the localized-magnon excitations
that we discussed earlier.

\begin{figure}[tb!]
\centering
\includegraphics[width=10 true cm]{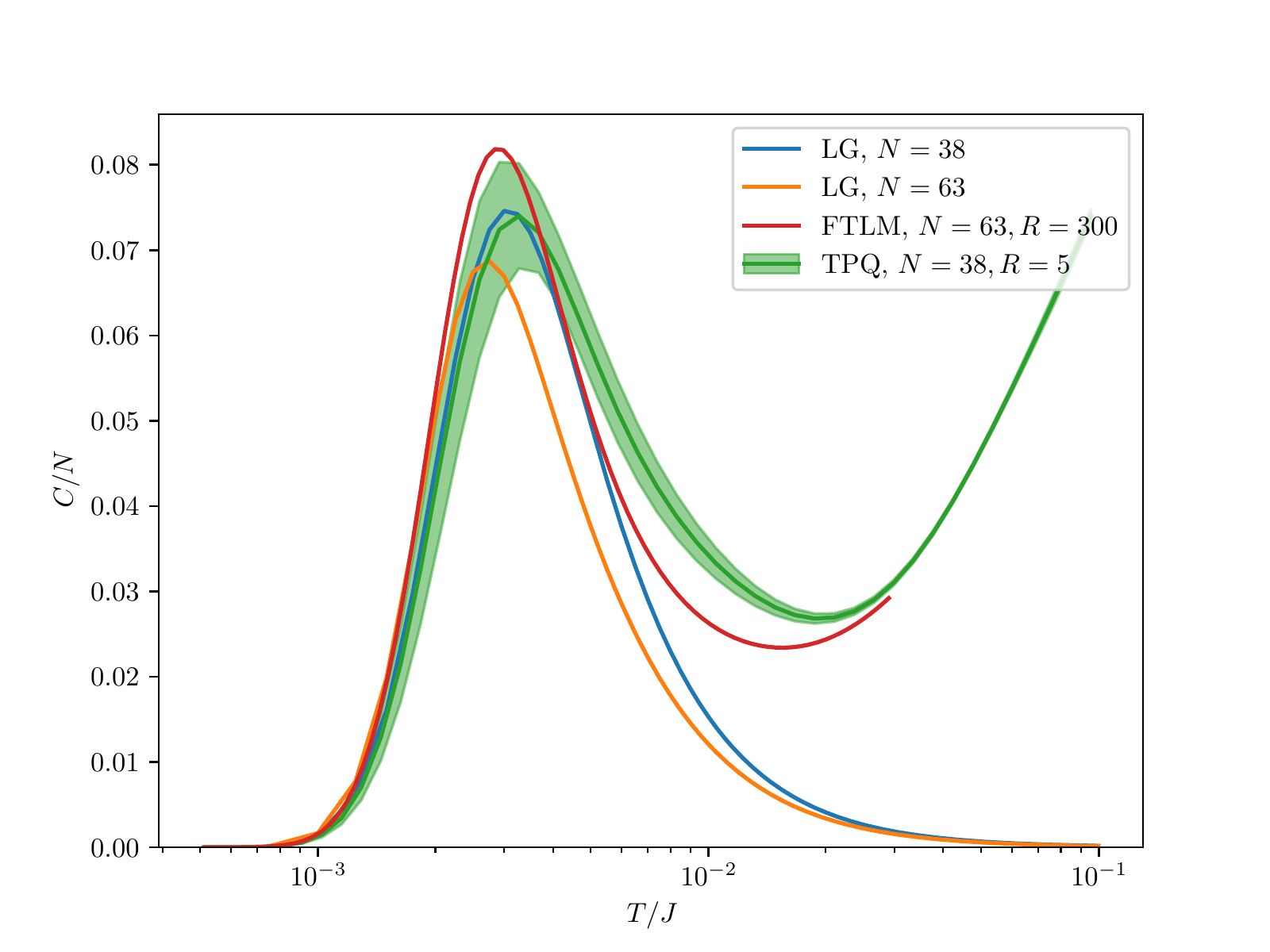} 
\caption{(Colour online) Low-temperature part of the specific heat per site $C/N$ at a 
magnetic field $h=0.99\,J$. Note the logarithmic temperature axis. Here, we 
compare results for the loop gas (LG) with TPQ for $N=38$ and FTLM for 
$N=63$, respectively.
\label{fig:specheat_loopgas}
}
\end{figure}

Figure \ref{fig:specheat_loopgas} presents a comparison of $N=38$ and $63$ 
data for the loop gas from figure~\ref{fig:ChhLoop} and for TPQ, 
respectively FTLM from figure~\ref{fig:specheat_edsizes}. First, we 
observe that for $N=38$, the loop gas matches TPQ for the spin-1/2 XY 
model exactly up to the low-temperature maximum of the specific heat $C$ 
around $T_{\max} \approx 0.3\cdot10^{-3}\,J$. This underlines again that 
the loop gas provides an exact description of the ground states of the 
spin-1/2 XY model in the regime of high magnetic fields. However, TPQ 
diverges from the loop gas at temperatures already immediately above 
$T_{\max}$. This can be traced to excitations that are not described by 
the loop gas appearing already at energies of the order of $10^{-2}\,J$, 
compare the values of the gap in table \ref{tab:3x3hex}. Likewise, the 
loop gas yields at least a qualitative description of the $N=63$ FTLM 
low-temperature peak in $C$. In this case, deviations are already visible 
around and below the temperature $T_{\max} \approx 0.3\cdot10^{-3}\,J$ of 
the maximum of $C$. We note that for $N=63$, the gap to excitations of the 
spin-1/2 XY model that are not captured by the loop gas has decreased to 
the order of $4\cdot10^{-3}\,J$, see table \ref{tab:4x4hex}, thus giving 
rise to further fluctuations already at quite low temperatures. 
Nevertheless, we may conclude that the low-temperature maximum of 
figure~\ref{fig:specheat_edsizes} corresponds to the localized-magnon 
states. In the case of the subclass of hard-hexagon states, this 
low-temperature maximum is known to correspond to a finite-temperature 
phase transition into the magnon crystal, see figure~\ref{fig:ChhLoop}. We 
expect the same to be true both for the loop gas and the spin-1/2 XY model 
even if the system sizes $N$ that would be required to exhibit the 
divergence in the specific heat $C$ are well beyond those accessible to us 
in the latter cases.

\section{Conclusion}

In this contribution, we have first reviewed the localized-magnon states 
that appear as ground states in the high-field regime of the spin-$s$ XXZ 
model on the kagome lattice 
\cite{SHS:PRL02,RDS:PRL04,ZhT:PRB04,ZhT:PTPS05,DeR:PRB04,DeR:EPJB06,DRH:LTP07}. 
A subset of these states is well understood in terms of hard hexagons 
\cite{ZhT:PRB04,ZhT:PTPS05}. However, it is also known that nested objects 
\cite{ZhT:PRB04,ZhT:PTPS05} yield another macroscopic contribution to the 
ground-state manifold \cite{DRH:LTP07}. Consequently, one needs to 
consider a more general loop-gas description. Since a previous 
investigation of finite kagome lattices \cite{SSHR20} demonstrated the 
existence of further ground states of the spin-1/2 Heisenberg model on 
lattices with periodic boundary conditions that are not captured even by 
the loop gas, we decided to investigate open boundary conditions in the 
present work in order to assess the completeness of the loop-gas 
description.

Indeed, here we found that the loop gas provides an exact match of the 
ground states of the spin-1/2 antiferromagnetic XY model on the kagome 
lattice in the corresponding sectors of $S^z$ when we impose open boundary 
conditions. Therefore, the loop gas should yield an exact description of 
the thermodynamic limit of the low-temperature behavior of the spin-1/2 
antiferromagnet at high magnetic fields independently of the boundary 
conditions. On the other hand, we also observed that open boundary 
conditions enhance finite-size effects, as might have been expected. One 
manifestation is that the lattices with $N\leqslant 72$ that were accessible to 
the present study turn out to be too small to accommodate a significant 
number of composite loop objects that cannot be mapped to hard hexagons 
such that one would have to go to bigger $N$ to clearly exhibit the 
macroscopic nature of the difference between loop-gas and hard-hexagon 
states on lattices with open boundary conditions.

Nevertheless, given that the loop gas should provide an accurate 
description of the ground-state manifold of the high-field regime of the 
spin-$s$ XXZ model on the kagome lattice, it would certainly be desirable 
to study this loop gas on bigger systems than we were able to do so far, 
or maybe even carry out the thermodynamic limit, as has been done for the 
subset of hard hexagons \cite{Baxter1980,BaxterTsang1980}, and thus also 
exhibit the crystallization transition in the loop gas. In this context, 
it would be helpful if the Gram-Schmidt orthogonalization that we have 
used to implement the linear relations among the states corresponding to 
multi-loop configurations could be avoided and a purely geometric 
description of the relevant loop configurations be provided instead.

A final chapter was devoted to the contribution of the localized-magnon 
states to low-temperature thermodynamic properties and a comparison with 
numerical results for the specific heat of the spin-1/2 XY model computed 
by exact-diagonalization variants such as thermal pure quantum (TPQ) 
states and the finite-temperature Lanczos method (FTLM). We showed that 
the macroscopic number of localized-magnon states gives rise to a 
low-temperature peak in the specific heat that in the thermodynamic limit 
develops into a finite-temperature phase transition. The low-temperature 
phase that arises in a finite window just below the saturation field 
corresponds to a magnon crystal that is formed at low temperatures by the
densest packing of the corresponding loop configurations.

In the present work, we have focused on the kagome lattice, but there are 
other lattices where a similar analysis of a loop-gas description of 
localized-magnons states could be performed, specifically the checkerboard 
lattice \cite{Richter2004,DRH:LTP07,ZT07} and the star lattice 
\cite{RSHS04} in two dimensions as well as the pyrochlore lattice 
\cite{ZT07} in three dimensions.

\section*{Acknowledgements}

We are indebted to O. Derzhko for his interest in and long-time 
collaborations on localized magnons. We would further like to thank M.E.~Zhitomirsky for providing us the data for the $N=\infty$ hard-hexagon 
curve in figure~\ref{fig:ChhLoop}. The Flatiron Institute is a division of 
the Simons Foundation. J.R.\ and J.S.\ thank the Deutsche 
Forschungsgemeinschaft for financial support (DFG grants RI 615/25-1 and 
SCHN 615/28-1). Part of the computations were performed on the osaka 
cluster at CY Cergy Paris Universit\'e.



\ukrainianpart

\title{Опис локалізованих магнонних станів на ґратці кагоме з відкритими граничними умовами за допомогою  петлевого газу}
%
\author{А. Гонекер \refaddr{adr:cergy}, Й. Ріхтер \refaddr{adr:magdeburg,adr:dresden}, Ю. Шнак \refaddr{adr:bielefld}, А. Вітек \refaddr{adr:flatiron}}
\addresses{
\addr{adr:cergy} Лабораторія теоретичної фізики та моделювання, CNRS UMR 8089, Університет Сержі Париж, 95302 Сержі-Понтуаз, Франція 
\addr{adr:magdeburg} Інститут фізики, Університет Магдебурга, поштова скринька 4120, Магдебург 39016, Німеччина
\addr{adr:dresden} Інститут Макса Планка фізики складних систем, Ньотнітцерштрасе 38, D--01187 Дрезден, Німеччина
\addr{adr:bielefld} Університет Білєфельда, факультет фізики, поштова скринька 100131, D--33501 Білєфельд, Німеччина
\addr{adr:flatiron} Центр обчислювальної квантової фізики, Флатиронський інститут, Нью-Йорк 10010, США
}

\makeukrtitle

\begin{abstract}
Режим високих полів у спін-$s$ XXZ антиферомагнетику на ґратці кагоме призводить до макроскопічно вироджених основних станів завдяки повністю плоскій найнижчій одномагнонній зоні. Відповідні збудження можуть бути локалізовані на петлях у реальному просторі і отримали назву ``локалізовані магнони''. Опис багаточастинкових основних станів зводиться до опису конфігурацій дозволених класичних петель та виключення квантово-механічних лінійних співвідношень між ними. Тут ми досліджуємо таке зображення петлевого газу на скінченних гратках кагоме з відкритими граничними умовами і порівнюємо результати точної діагоналізації з спін-1/2 XY моделлю на такій самій ґратці. Ми знайшли, що петлевий газ забезпечує точне врахування многовиду вироджених станів, в той час як опис жорсткими гексагонами пропускає вклади конфігурацій вкладених петель. Найщільніше пакування петель відповідає магнонному кристалу, що згідно  з кривою намагніченості при нульовій температурі, є стабільним основним станом спін-1/2 XY моделі у діапазоні магнітних полів біля $4\%$ нижче поля насичення. Ми також представляємо числові результати для теплоємності, отримані за допомогою пов'заних методів термічних чистих квантових станів та скінченно-температурного методу Ланцоша. Для поля в області стабільності магнонного кристалу, знайдено низькотемпературний максимум теплоємності, що відповідає скінченно-температурному фазовому переходу до магнонного кристалу при низьких температурах.

\keywords фрустрований магнетизм, гратка кагоме, XY модель, граткові гази, фазові переходи, точна діагоналізація
\end{abstract}
 
\end{document}